\documentclass[reprint,prd,nofootinbib,amsmath,amssymb,aps,twocolumn,superscriptaddress]{revtex4-2}
\usepackage{hyperref}
\usepackage{graphicx}	% Including figure files
\usepackage{bm}
\usepackage[utf8]{inputenc}
\usepackage{xcolor}
\usepackage{multirow}

\usepackage{natbib}
\defcitealias{2023ApJ...946..106C}{C23}
\defcitealias{2022ApJ...928...65H}{HC22}
\defcitealias{2023JCAP...01..044D}{DP23}

\def\addtodot#1.#2\relax{#1\rlap{.}^{\dotadd}#2}
\newcommand{\dotdeg}[1]{\def\dotadd{\circ}\addtodot#1\relax}
\newcommand{\dotarcmin}[1]{\def\dotadd{\prime}\addtodot#1\relax}
\newcommand{\vect}[1]{\ensuremath{\bm{#1}}}
\newcommand{\matr}[1]{\ensuremath{\bm{\mathrm{#1}}}}
\DeclareMathOperator{\atantwo}{atan2}

\begin{document}

%\preprint{}

%\title{Filaments can model/explain/reproduce parity violating spectra in Galactic thermal dust: implications for cosmic birefringence}
%\title{Filaments can model parity violating spectra in Galactic thermal dust: implications for cosmic birefringence}
\title{Modeling parity-violating spectra in Galactic dust polarization with filaments and its applications to cosmic birefringence searches}


\author{Carlos Herv\'ias-Caimapo}
\affiliation{Instituto de Astrof\'isica and Centro de Astro-Ingenier\'ia, Facultad de F\'isica, Pontificia Universidad Cat\'olica de Chile, Av. Vicu\~na Mackenna 4860, 7820436 Macul, Santiago, Chile}
\email{carlos.hervias@uc.cl}
\author{Ari J. Cukierman}
\affiliation{Department of Physics, California Institute of Technology, Pasadena, CA 91125, USA}
\author{Patricia Diego-Palazuelos}
\affiliation{Max-Planck-Institut f\"{u}r Astrophysik, Karl-Schwarzschild Str. 1, 85741 Garching, Germany}
\author{Kevin M. Huffenberger}
\affiliation{Department of Physics, Florida State University, Tallahassee, Florida 32306, USA}
\affiliation{Mitchell Institute for Fundamental Physics \& Astronomy and  Department of Physics \& Astronomy, Texas A\&M University, College Station, Texas 77843, USA}
\author{Susan E. Clark}
\affiliation{Department of Physics, Stanford University, Stanford, CA 94305, USA}
\affiliation{Kavli Institute for Particle Astrophysics \& Cosmology, P.O. Box 2450, Stanford University, Stanford, CA 94305, USA}


\date{\today}

\begin{abstract}
We extend the dust-filament-based model presented in Herv\'ias-Caimapo \& Huffenberger 2022 to produce parity-violating foreground spectra by manipulating the filament orientations relative to the magnetic field. We calibrate our model to observations of the misalignment angle using cross-correlations of \textit{Planck} and HI 21-cm line data, producing a fiducial model that predicts a $\mathcal{D}_{\ell}^{EB}\sim$few\,$\mu$K$^2$ dust signal at 353\,GHz and where $\sim 56$\% of filaments have a positive misalignment angle. The main purpose of this model is to be used as dust with non-zero parity-violating emission in forecasting a measurement of cosmic birefringence by upcoming experiments. Here, we also use our fiducial model to assess the impact of dust in measurements of the isotropic cosmic birefringence angle $\beta$ with \textit{Planck} data by measuring the misalignment angle as a function of scale, as well as directly using our model's $\mathcal{D}_{\ell}^{EB}$ prediction as a template. In both cases, we measure $\beta$ to be consistent within $0.83\sigma$ of the analyses that use the 353\,GHz channel of \textit{Planck} to constrain the filamentary misalignment or that use the \textsc{commander} sky model as a template for the dust $\mathcal{D}_{\ell}^{EB}$ spectrum. We attribute this consistency to the extra degrees of freedom introduced by the ad-hoc amplitude parameters for the dust parity-violating spectrum, which are capable of absorbing most of the differences between the model and data as long as a reasonable dust model is provided.
\end{abstract}

%\keywords{}

\maketitle

%\tableofcontents

\section{Motivation} \label{sec:intro}

Millimeter emission from diffuse Galactic foregrounds is one of the most important sources of contamination for the observation of cosmic microwave background~(CMB) radiation, especially in polarization. Of these, diffuse thermal emission from dust and synchrotron radiation are the main contributions~\cite{2016A&A...594A..10P,2020A&A...641A...4P}. They impact cosmological measurements of the late universe such as the gravitational lensing of CMB photons by the intervening large-scale structure~\cite[e.g.][]{2012JCAP...12..017F,2020JCAP...06..030B,2022MNRAS.514.5786B}, as well as the yet-to-be-detected large-scale polarized signal predicted by the production of a hypothetical stochastic background of gravitational waves~\cite{2016ARA&A..54..227K} or even a hypothetical primordial non-Gaussianity~\cite[e.g.][]{2010MNRAS.405..961C,2018JCAP...11..047J,2019JCAP...10..056C,2020A&A...641A...9P}, both sourced in the very birth of the Universe at high energy ranges unobtainable anywhere else experimentally. However, in this work, we focus on the potential impact of polarized foregrounds on physics beyond the standard model of cosmology involving parity violation. While no intrinsic parity violation has been detected so far from synchrotron~\cite{2022JCAP...04..003M,2023MNRAS.519.3383R}, thermal dust does have a parity-violating spectrum measured by \textit{Planck}~\cite{2020A&A...641A..11P}.

% introduce E/B modes and parity violation
The CMB radiation is linearly polarized. Its Stokes parameters $Q$ and $U$ can be combined into $E$ and $B$ fields~\cite{1997PhRvL..78.2058K,1997PhRvL..78.2054S,1997PhRvD..55.1830Z}. Under an inversion of spatial coordinates, the $E$ field is parity even, while the $B$ field is parity odd. The angular (auto) power spectra we define from these fields, i.e., the $C_{\ell}^{EE}$ and $C_{\ell}^{BB}$ spectra, are invariant under parity transformation, while the $C_{\ell}^{EB}$ cross-power spectrum changes sign under parity transformation. Therefore, the CMB radiation is sensitive to parity violation through the $C_{\ell}^{EB}$ spectrum~\cite{1999PhRvL..83.1506L}. Since the intensity field $T$, just like $E$, is parity even, then the $C_{\ell}^{TB}$ cross-power spectrum would also be sensitive to parity violation.

% introduce cosmic birefringence
An unknown parity-breaking mechanism or interaction acting on traveling CMB photons could imprint a measurable signature. An example of such a phenomenon is an axion-like pseudo-scalar field that couples to the electromagnetic tensor via a Cherns-Simons term in the Lagrangian density~\cite{1983PhRvL..51.1415S,1988PhRvD..37.2743T}. Under the assumption of spatial homogeneity, if the pseudo-scalar field slowly evolves with time, e.g., like a quintessence field, the plane of linear polarization of photons will rotate by an angle $\beta$~\cite{1990PhRvD..41.1231C,1992PhLB..289...67H,1998PhRvL..81.3067C}. This rotation is denominated ``cosmic birefringence'', in analogy to the universe being filled with a birefringent fluid in which circular polarization states propagate at different velocities producing a net rotation. See Ref.~\cite{2022NatRP...4..452K} for a review. Models have been proposed where an axion-like field is a candidate for both dark matter and dark energy~\cite{2010ARA&A..48..495F,2016PhR...643....1M,2021A&ARv..29....7F}, so a detection of cosmic birefringence would profoundly impact our understanding of the nature of the Universe.

% current measurements of CB
% here maybe we can talk about measuring alpha+beta until you have a way to break the degeneracy ?
In the last few years, hints of a possible detection of cosmic birefringence have been measured. Ref.~\cite{2020PhRvL.125v1301M} first presented a measurement of $\beta = \dotdeg{0.35} \pm \dotdeg{0.14}$, a $2.4\sigma$ measurement, using \textit{Planck} High Frequency Instrument~(HFI) 2018 data~\cite{2020A&A...641A...3P}. This method exploits the observation of the CMB together with Galactic foregrounds to break the degeneracy between an instrumental polarization angle and a proper cosmological birefringence angle~\cite{2019PTEP.2019h3E02M,2020PTEP.2020j3E02M}. Subsequent works have included more data as well as refined the method~\cite{2022PhRvL.128i1302D,2022A&A...662A..10E,2022PhRvD.106f3503E,2023A&A...679A.144E}. Ref.~\cite{2022PhRvD.106f3503E} presents the tightest constraints of the cosmic birefringence angle to date, $\beta = \dotdeg{0.342} \substack{+\dotdeg{0.094}\\-\dotdeg{0.091}}$, a $\sim 3.6 \sigma$ measurement, using the \textit{Planck} \textsc{npipe} maps~\cite{2020A&A...643A..42P} over nearly the full sky, together with the WMAP 9-year observations~\cite{2013ApJS..208...20B}. However, any intrinsic non-zero parity-violating spectra from local foregrounds must be accounted for. While these works consider a $C_\ell^{EB}$ signal from foregrounds in one way or another, more effort is needed to understand their impact fully: from better modeling of the dust spectral energy distribution (SED)~\cite{2023A&A...672A.146V, Vacher2023, Ritacco2023, Vacher2024} and derivation of a signal-dominated dust template~\cite{2023JCAP...01..044D} to a better understanding of the physical mechanism generating the parity-violating $TB$ and $EB$ signals~\cite{2020ApJ...899...31H,2021ApJ...919...53C} and the construction of robust estimators of the magnetic misalignment angle~\cite{2023ApJ...946..106C,2024ApJ...961...29H}.

% Planck has measured TB > 0
The \textit{Planck} mission has measured a positive $TB$ power spectrum in the 353\,GHz frequency channel, dominated by thermal dust emission, while the $EB$ spectrum is consistent with zero~\cite{2016A&A...586A.133P,2020A&A...641A..11P}. The ratio between a power-law fit of the $TB$ and $TE$ spectra in the multipole range $\ell = 40-600$ is $\sim 0.1$ (with an anchor angular scale of $\ell=80$), which would translate to an amplitude $A^{TB}(\ell=80) \sim 80$\,$\mu$K$^2$ for the largest sky fraction ($\sim 71$\%) considered in these works. Further analysis correlating \textit{Planck} 353\,GHz observations with independent data, such as lower frequency channels from WMAP dominated by synchrotron or optical polarized starlight, also find a positive $TB$ spectrum~\cite{2020ApJ...893..119W}. In the diffuse emission from our Galaxy, the polarization of dust is the product of the interplay of elongated dust grains aligned with respect to the Galactic magnetic field~\cite{2003ARA&A..41..241D}. Synchrotron\footnote{Synchrotron radiation is weakly correlated to thermal dust emission~\cite{2015JCAP...12..020C}, despite the former probing a larger path length~\cite{2024ApJ...966...43M}, and its emission depending on both the Galactic magnetic field strength and local cosmic ray properties~\cite{2025ApJ...980..197D}.} and polarized starlight are independent tracers of the magnetic field, so this analysis supports the idea that a positive $TB$ spectrum from dust is a real feature in the millimeter emission from our Galaxy. 

% introduce filaments
Interstellar dust grains tend to align their short axes parallel to the local magnetic field, which induces a coherent polarized emission~\cite{1975duun.book..155P,2003ARA&A..41..241D}. The morphology of diffuse Galactic dust seems to be partially composed of a filamentary structure~\cite{2023ASPC..534..153H}. These filaments have been previously observed and characterized in the millimeter~\cite{2016A&A...586A.136P,2016A&A...586A.141P} as well as other wavelengths (e.g.~\cite{2009ApJ...700.1609M,2010A&A...518L.100M}). Moreover, Galactic emission from the 21-cm hyperfine transition from neutral hydrogen is strongly correlated to dust~\cite{1996A&A...312..256B,2017ApJ...846...38L}, which enables the study of dust in a third dimension along the line of sight through the Doppler shift of different velocity components. The filaments seen in HI are well aligned with the local interstellar magnetic field being traced either by starlight polarization~\cite{2006ApJ...652.1339M,2014ApJ...789...82C} or by dust millimeter emission~\cite{2015PhRvL.115x1302C,2015ApJ...809..153M,2023ApJ...945...72A}. Furthermore, HI can be used to predict what the dust millimeter polarized emission will look like~\cite{2019ApJ...887..136C}.

% filaments could explain != 0 parity violation
Dust filaments have been invoked as one possible explanation for the non-zero parity-violating $TB$ spectrum. Ref.~\cite{2020ApJ...899...31H} put forward the idea that a certain degree of misalignment between the filaments and the magnetic field can quantitatively describe the statistical properties of Galactic dust as seen by \textit{Planck} in Ref.~\cite{2020A&A...641A..11P}, as well as parity-violating spectra by appealing to an asymmetry in the handedness of this misalignment angle, e.g., having more filaments with a positive misalignment angle than a negative one. Furthermore, Ref.~\cite{2021ApJ...919...53C} presented evidence that the dust positive $TB$ is driven by a coherent misalignment between the dust ISM filaments and the magnetic field projected onto the plane of the sky. This misalignment angle, labeled $\psi$, was also measured to be roughly scale independent, with a value $\psi \sim 5^{\circ}$ in the multipole range $100 \lesssim \ell \lesssim 500$. As the follow up of the previous work, Ref.~\cite{2023ApJ...946..106C}, hereafter \citetalias{2023ApJ...946..106C}, refined the analysis by defining new estimators for the angle $\psi$, finding a robust $\psi \sim (2 \pm 1)^{\circ}$ scale-independent value in the multipole range $100 \lesssim \ell \lesssim 700$.

% Other work on parity violation spectra from Galactic foregrounds besides filaments
Other works have tried to explain the non-zero parity-violating dust $TB$ spectrum by invoking features in the interstellar magnetic field and the magnetohydrodynamic (MHD) turbulence. For example, Ref.~\cite{2019A&A...621A..97B} produces non-zero $TE$ and $TB$ correlations at $\ell \lesssim 20$ scales by invoking magnetic helicity~\cite{2015SSRv..188...59B,2020MNRAS.499.3673W} in the local solar neighborhood. The notions of an asymmetry in the filament-magnetic field misalignment and a helicity in our local volume are complementary. Ref.~\cite{2022Univ....8..423H} finds the \textit{Planck}-observed dust $TB$ spectrum is inconsistent with a pure statistical fluctuation of filament misalignment. Given this, there must be an underlying physical mechanism for the preference of the filaments' magnetic misalignment.

% Implications of a parity-violating spectra for cosmic birefringence. Remember to mention we would expect EB > 0. Mention how EB_dust was handled in the latest CB papers.
Given all of the evidence for positive $TB$ correlation from thermal dust emission, in this filament misalignment model we would expect the $EB$ correlation also to be positive (even if \textit{Planck} does not have enough sensitivity to detect it) and therefore to significantly impact measurements of cosmic birefringence using the method pioneered in Ref.~\cite{2020PhRvL.125v1301M}. Two approaches to account for a potential non-zero dust $EB$ spectrum were introduced: one is using a template of thermal dust to directly estimate the $EB$ spectrum from maps~\cite{2022PhRvL.128i1302D,2023JCAP...01..044D}, and the other, used in Refs.~\cite{2022PhRvL.128i1302D,2022A&A...662A..10E,2022PhRvD.106f3503E}, is to adopt the magnetic misalignment of filaments ansatz presented in Ref.~\cite{2021ApJ...919...53C} and assume that the dust $EB$ is proportional to dust $TB$, which leads to
\begin{equation} \label{eq:EB_from_dust}
    C_{\ell}^{EB,\rm d} = A_{\ell} C_{\ell}^{EE,\rm d} \sin(4 \psi_{\ell}) \text{,}
\end{equation}
where $A_{\ell}$ is a free amplitude, and $\psi_{\ell}$ is the scale-dependent misalignment angle estimated from the dust spectra,
\begin{equation} \label{eq:psi_TB_TE}
    \psi_{\ell} = \frac{1}{2} \arctan \left( \frac{C_{\ell}^{TB,\rm d}}{C_{\ell}^{TE,\rm d}} \right) \text{.}
\end{equation}
The spectra-based estimator of eq.~\eqref{eq:psi_TB_TE} depends only on dust observations from \textit{Planck}, which likely includes contributions from non-filamentary dust emission, as well as systematics, potentially distorting the measurement. Alternatively, following \citetalias{2023ApJ...946..106C}, we will explore the use of HI data as a tracer of filaments and of different $\psi_{\ell}$ estimators in cosmic birefringence analysis, among other aspects of the impact of parity-violating dust.

% Introduce the Hervias-caimapo 2022 model and how we can easily inject assymetry
The \textsc{dustfilaments} model presented in Ref.~\cite{2022ApJ...928...65H}, hereafter \citetalias{2022ApJ...928...65H}, simulates an actual realization of a population of millions of filaments in a cubic volume, projecting the view into an observer located at the center to produce a full-sky map of intensity and polarization of the millimeter emission of dust. In~\citetalias{2022ApJ...928...65H}, filaments are oriented randomly with respect to the underlying magnetic field, so no asymmetry in the handedness of $\psi$ is produced deliberately, and the $TB$ and $EB$ correlations are therefore consistent with zero. However, a natural extension of this model is to force filaments to show an asymmetry in the handedness of $\psi$ by design, producing non-zero parity-violating spectra in the process. \citetalias{2022ApJ...928...65H} used the $EE$, $BB$, and $TE$ \textit{Planck} spectra to constrain the filament model, but refrained from modeling parity-violating correlations since \textit{Planck} $TB$ and $EB$ spectra are not sensitive enough on their own to constrain a model that accounts for the filament asymmetry. The goal of this paper is to produce a realistic simulation of the millimeter Galactic dust that includes sensible non-zero $TB$ and $EB$ spectra. We resort to calibrating our model using \textit{Planck} observations, as well as external data in the form of HI surveys tracing the filament structure, following~\citetalias{2023ApJ...946..106C}. This model can then be used for forecasting the impact of parity-violating dust in cosmology in the context of future CMB experiments, as well as be applied to current measurements of cosmic birefringence.

% Outline of the paper.
Our paper is organized as follows. Section~\ref{sec:data} details the \textit{Planck} and HI data we use throughout this work. Section~\ref{sec:filament_model} summarizes the filament model presented by~\citetalias{2022ApJ...928...65H}, as well as the mechanism for achieving an asymmetry in the misalignment. Section~\ref{sec:psi_estimators} introduces the estimators for the misalignment angle $\psi$ and how they are measured from the cross-correlation between \textit{Planck} and HI data. Section~\ref{sec:results} details how we fit our model to observations and presents the results for the dust model producing parity-violating spectra, including our fiducial model showing a prediction for the Galactic dust $EB$ spectrum.
Section~\ref{sec:cosmic_birefringence} presents applications of our model to measurements of cosmic birefringence, analyzing the impact of parity-violating dust. Section~\ref{sec:discussion} discusses how our Galaxy could have the apparent asymmetry in the filament misalignment physically. Finally, in Section~\ref{sec:conclusions} we summarize and present our conclusions.
\section{Data} \label{sec:data}

Our main source of thermal dust observations is the \textit{Planck} mission\footnote{All data products are available at the \textit{Planck} Legacy Archive \url{https://pla.esac.esa.int}.}~\cite{2020A&A...641A...1P} and its polarized HFI 353\,GHz channel. As a tracer of filaments, we also use HI full-sky spectra from the HI4PI survey~\citep{2016A&A...594A.116H}.

\subsection{\textit{Planck} frequency maps and dust models} \label{sec:planck_data}

Like~\citetalias{2023ApJ...946..106C}, we use the \textsc{commander} dust maps estimated with parametric component separation~\cite{2008ApJ...676...10E} to isolate the dust emission. We use the full mission \textsc{commander} map constructed with all the available data, as well as two half-mission maps constructed from either the first or second half of the observing run when we have to calculate cross-power spectra and we want to avoid noise bias.

For masking, we use the \textit{Planck} Galactic plane masks from Public Release~(PR) 2. In particular, we use the mask with sky fraction $f_{\rm sky}=70$\% as used in~\citetalias{2023ApJ...946..106C} with an apodization scale of $1^{\circ}$. The \textsc{commander} dust maps are also smoothed with a Gaussian beam with a full width at half maximum (FWHM) of $\dotarcmin{16.2}$, which is the resolution of the HI4PI survey.

\begin{table}
    \caption{Fitted parameters of the power-law model, eq.~\eqref{eq:power-law}, to the dust power spectra estimated from the \textit{Planck} \textsc{npipe} 353\,GHz frequency map in the Galactic 70\% mask. For $TB$, we fix the power-law index to $\alpha=-2.44$ and only fit the amplitude $A$. \label{table:dust_spectra_npipe}}
    \begin{ruledtabular}
    \begin{tabular}{c|c|c|c}
        Spectrum & $\ell$ range & $A$ [$\mu$K$^2$] & $\alpha$ \\
        \hline
        $TT$ & $260-600$ & $17,035 \pm 1,305$ & $-2.46 \pm 0.05$ \\
        $TE$ & $40-600$ & $598.6 \pm 23.3$ & $-2.44 \pm 0.04$ \\
        $EE$ & $40-600$ & $203.6 \pm 3.7$ & $-2.40 \pm 0.03$ \\
        $BB$ & $40-600$ & $121.9 \pm 1.7$ & $-2.55 \pm 0.03$ \\
        $TB$ & $40-600$ & $39.4 \pm 6.4$ & $-2.44$ \\
    \end{tabular}
    \end{ruledtabular}
\end{table}

Regarding our filament model, we re-estimate the power spectra from Galactic dust for our particular needs in this study. In~\citetalias{2022ApJ...928...65H}, we calibrated the filament model to the spectra estimated with the PR3 353\,GHz map~\cite{2020A&A...641A..11P} in the Large Region (LR) 71 mask~\cite{2016A&A...586A.133P}. In this work, we re-calculate the dust spectra in the same way as done in Ref.~\cite{2020A&A...641A..11P} but using the Galactic plane 70\% mask, as well as updating the 353\,GHz frequency map to the latest \textsc{npipe} maps~\cite{2020A&A...643A..42P} instead of using PR3. Also, we add the masking of strong polarized point sources from Ref.~\cite{2016A&A...586A.133P} to the Galactic 70\% mask to estimate the dust power spectra, since they can bias the high-$\ell$ spectrum. For cross-spectra, we use the A and B detector splits to avoid a noise bias. Using the same binning scheme as Table C.1 from Ref.~\cite{2020A&A...641A..11P}, we fit the following power law model to each spectrum
\begin{equation} \label{eq:power-law}
    \mathcal{D}_{\ell}^{XY} = A^{XY} (\ell / 80)^{\alpha_{XY}+2} \text{,}
\end{equation}
where $XY \in [TT, TE, EE, BB]$ and $\mathcal{D}_{\ell} \equiv \frac{(\ell+1) \ell}{2 \pi} C_{\ell}$. We also estimate a power-law fit to the $TB$ spectrum of dust, but fixing $\alpha_{TB}=-2.44$ and only fitting for $A^{TB}$, following Ref.~\cite{2020A&A...641A..11P}, where the 353\,GHz frequency map does not have enough constraining power to fit both an amplitude and a power-law index. We estimate the spectra error bars from 200 realizations of the official end-to-end \textsc{npipe} simulations for the HFI 353\,GHz frequency channel\footnote{Available at NERSC at \path{/global/cfs/cdirs/cmb/data/planck2020/npipe}.}, including CMB, foregrounds, noise, and systematics. In Table~\ref{table:dust_spectra_npipe}, we summarize the power-law parameters for our fit. 

When creating a realization of our filament model, we use the \textsc{gnilc} $T$ dust map~\cite{2016A&A...596A.109P} with a fixed resolution of $80'$ as our template of the Galactic emission to place filaments in the celestial sphere, just like we did in~\citetalias{2022ApJ...928...65H}.

For all power spectra estimation required in this work, we compute spectra with the \textsc{namaster}\footnote{\url{https://github.com/LSSTDESC/NaMaster}} software~\cite{2019MNRAS.484.4127A}. When apodization is needed, we use the $C^2$ window.

\subsection{HI data, HI4PI} \label{sec:hi_data}
We use the HI4PI survey~\citep{2016A&A...594A.116H} and its full-sky observation of the 21-cm line at an angular resolution of $\dotarcmin{16.2}$ and at a spectral resolution of $\Delta v_{\rm lsr} = 1.49$\,kms$^{-1}$. This full-sky survey is achieved by combining the northern sky observed with the Effelsberg-Bonn HI survey~\cite{2016A&A...585A..41W} and the southern sky observed with the Parkes Galactic All-Sky Survey~\cite{2009ApJS..181..398M}. While HI4PI has a broad spectral width of $\pm$ hundreds of km\,s$^{-1}$, following~\citetalias{2023ApJ...946..106C}, only a few low-velocity bins are used in the range $-15$\,km\,s$^{-1} \leq v_{\rm lsr} \leq +4$\,km\,s$^{-1}$ \cite{2020ApJ...902..120P}.
\section{Filament model} \label{sec:filament_model}

In this section, we briefly summarize the thermal dust filament model presented in~\citetalias{2022ApJ...928...65H}. This will produce $TQU$ maps where filaments will have no preference on the handedness of the projected misalignment angle, $\psi$. Therefore, this baseline model produces $C_{\ell}^{TB}$ and $C_{\ell}^{EB}$ consistent with zero. Then, we will detail how we modify the model to produce a preference for the handedness of $\psi$ and therefore non-vanishing $C_{\ell}^{TB}$ and $C_{\ell}^{EB}$.

\subsection{Summary of filament model} \label{sec:summary}

\begin{figure}
    \includegraphics[width=1.0\columnwidth]{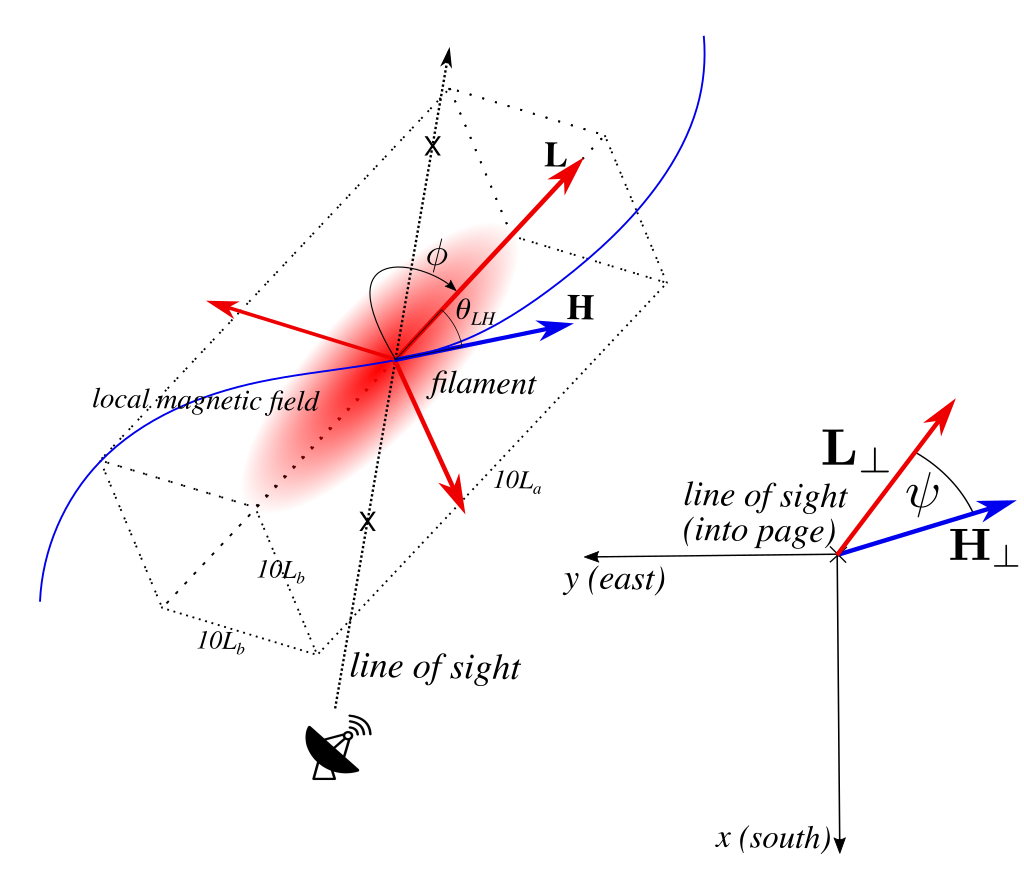}
    \caption{Diagram showing the relevant angles and the geometry for a single filament. $\theta_{\rm LH}$ is the angle between the filament and magnetic field in 3D space, while $\phi$ is the azimuthal random angle of the filament around the magnetic field. Finally, $\psi$ is the angle between the projections of the two vectors into the plane of the sky. While the filament is easily represented as a vector, it is truly a headless vector where a rotation has a period of $\pi$ (rotating by angle $\theta_{\rm LH}$ or $\theta_{\rm LH}+\pi$ is equivalent).}
    \label{fig:filament}
\end{figure}

First, we define a cubic volume of (400\,pc)$^3$ that is populated with a magnetic field with a resolution of $256^3$ voxels composed of a large-scale Galactic model~\citep{2012ApJ...757...14J,2012ApJ...761L..11J} and a random magnetic field $\vect{H}$ drawn from a power-law spectrum such that $\nabla \cdot \vect{H}=0$. Next, we place filaments at random positions inside. Individual filaments are modeled as prolate spheroids with a long semi-axis $L_a$ and two equal short semi-axes $L_b$. Filaments' sizes are defined by a power-law distribution $p(L_a) \propto L_a^{-\eta_L}$, and orientation angles are produced randomly. The filament long axis $\vect{L}$ is rotated with respect to the local magnetic field $\vect{H}$ by an angle $\theta_{\rm LH} \sim \mathcal{N}(\mu=0, \sigma^2=\text{rms}(\theta_{\rm LH})^2)$. Then, $\vect{L}$ is rotated around $\vect{H}$ by a random azimuthal angle $\phi \sim \mathcal{U}(0,2\pi)$. Both rotations are achieved using Rodrigues' rotation formula. The angle between $\vect{L}$ and $\vect{H}$, projected into the plane of the sky is $\psi$. The relevant geometry for a single filament is illustrated in Fig.~\ref{fig:filament}.

We repeat this procedure for many filaments, whose radial distance is determined by a random uniform distribution and whose azimuthal and polar angle coordinates can be fixed following a map template, e.g., the intensity from the \textit{Planck} \textsc{gnilc} dust map~\cite{2016A&A...596A.109P}, in order to reproduce the intensity pattern of the Galactic plane. We integrate along the line of sight from all filaments to an observer in the center of the box, producing a $TQU$ map. The physical size of the model places physical distances on where the filaments are placed (up to $\sim 160$\,pc from the observer). The only practical effect of putting a physical scale is to determine the large-scale magnetic field model of the Galactic disk. However, this is sub-dominant to the random power-law magnetic field. While we acknowledge most of the dust emission originates beyond the Local Bubble~\cite{2024ApJ...973...54H}, a physical scale has virtually no effect in the final integrated map.

\begin{figure}
    \includegraphics[width=1.0\columnwidth]{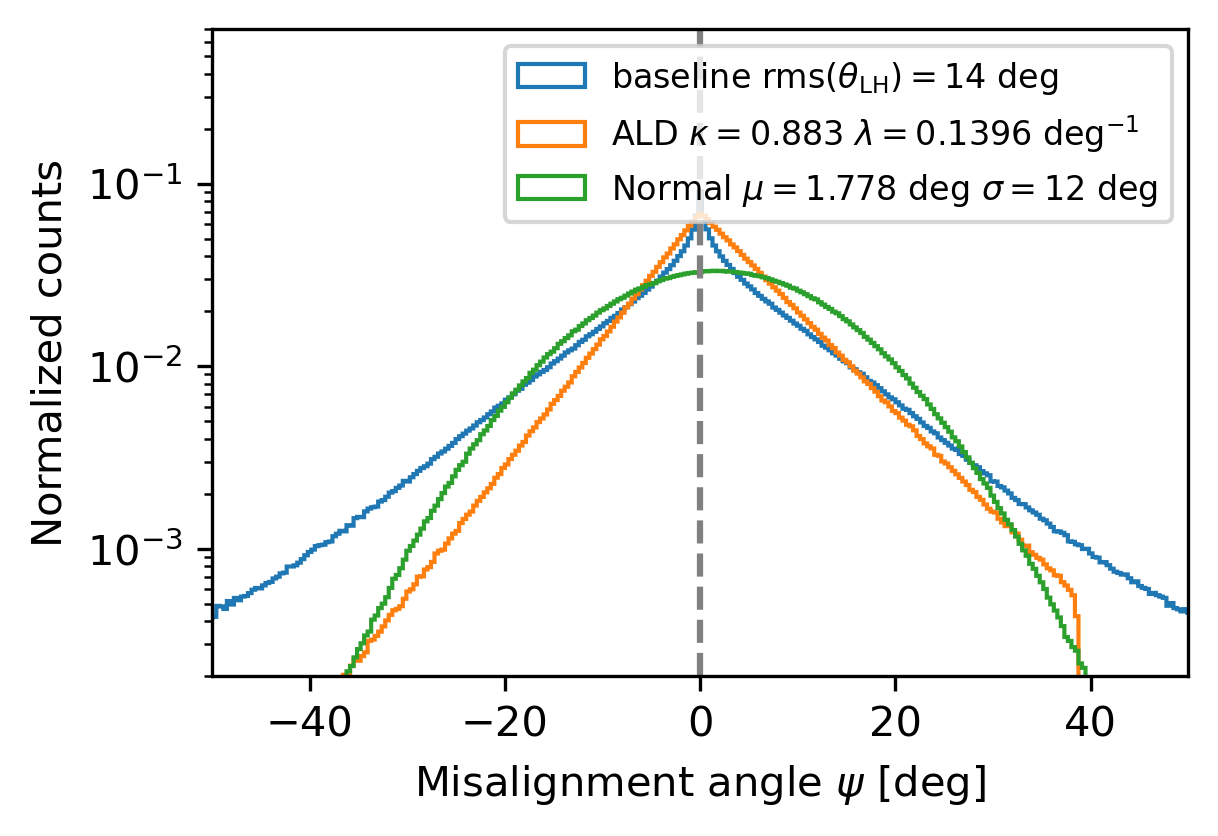}
    \caption{Histograms of measured $\psi$ angles for an example population of 10 million filaments. Note the ALD distribution and its wings with ranges of nonphysical angles $\psi \gtrsim 40^{\circ}$. \label{fig:psi_distribution}
    }
\end{figure}

The solid blue line in Fig.~\ref{fig:psi_distribution} shows the measured $\psi$ angles for an example population of 10 million filaments using the baseline model described above, with ${\rm rms}(\theta_{\rm LH})=14^{\circ}$. This distribution is symmetric around zero, and the parity-violating spectra simulated from such a distribution would be consistent with zero. We refer the reader to~\citetalias{2022ApJ...928...65H} for a detailed description of how the filament model works.

\subsection{Mechanism for asymmetric $\psi$} \label{sec:asymmetric_psi}

Here, we describe how the filament orientation is manipulated to achieve a particular distribution on the misalignment angle $\psi$ between the filament and the local magnetic field projected into the plane of the sky.

The orientation of a filament has three relevant angles: $\theta_{\rm LH}$, $\phi$, and $\psi$. Setting two of these three will fix the third angle into two values that are equivalent, meaning that the projection into the plane of the sky as seen by an observer looks exactly the same. For example, in the baseline model presented in~\citetalias{2022ApJ...928...65H}, $\theta_{\rm LH}$ and $\phi$ are fixed, and that sets the $\psi$ angle. To achieve the asymmetry in the distribution of $\psi$ angles, we randomly draw them from a predefined probability distribution. Therefore, we fix the $\theta_{\rm LH}$ and $\psi$ angles for each filament, and that will set two values for the $\phi$ angle that are equivalent.

From the geometry of the problem, we immediately note that $\theta_{\rm LH}$ and $\psi$ cannot be completely independent variables. $\theta_{\rm LH}$ represents the 3D angle between the long axis of a filament and the magnetic field. Given a fixed value of $\theta_{\rm LH}$, we note that the maximum angle between the filament and magnetic field \emph{projected into the plane of the sky} will be at most $\theta_{\rm LH}$. Therefore, we have the condition 
\begin{equation} \label{eq:module_condition}
    |\psi| \leq |\theta_{\rm LH}| \text{,}
\end{equation}
and any filament that does not meet this condition would be nonphysical.

We create two correlated angle random variables $\theta_{\rm LH}$ and $\psi$ using inverse transform sampling. We create a random variable $u \in \mathcal{U}(0,1)$. Let $X \in [\theta_{\rm LH}, \psi]$ be the random variable of the angles, with cumulative distribution function (CDF) $F_X$. The generalized inverse of the CDF evaluated with $u$, $X'(u) = F_X^{-1}(u)$ has distribution $F_X$ and therefore the same probability distribution as $X$.

\begin{figure*}
\includegraphics[width=0.49\textwidth]{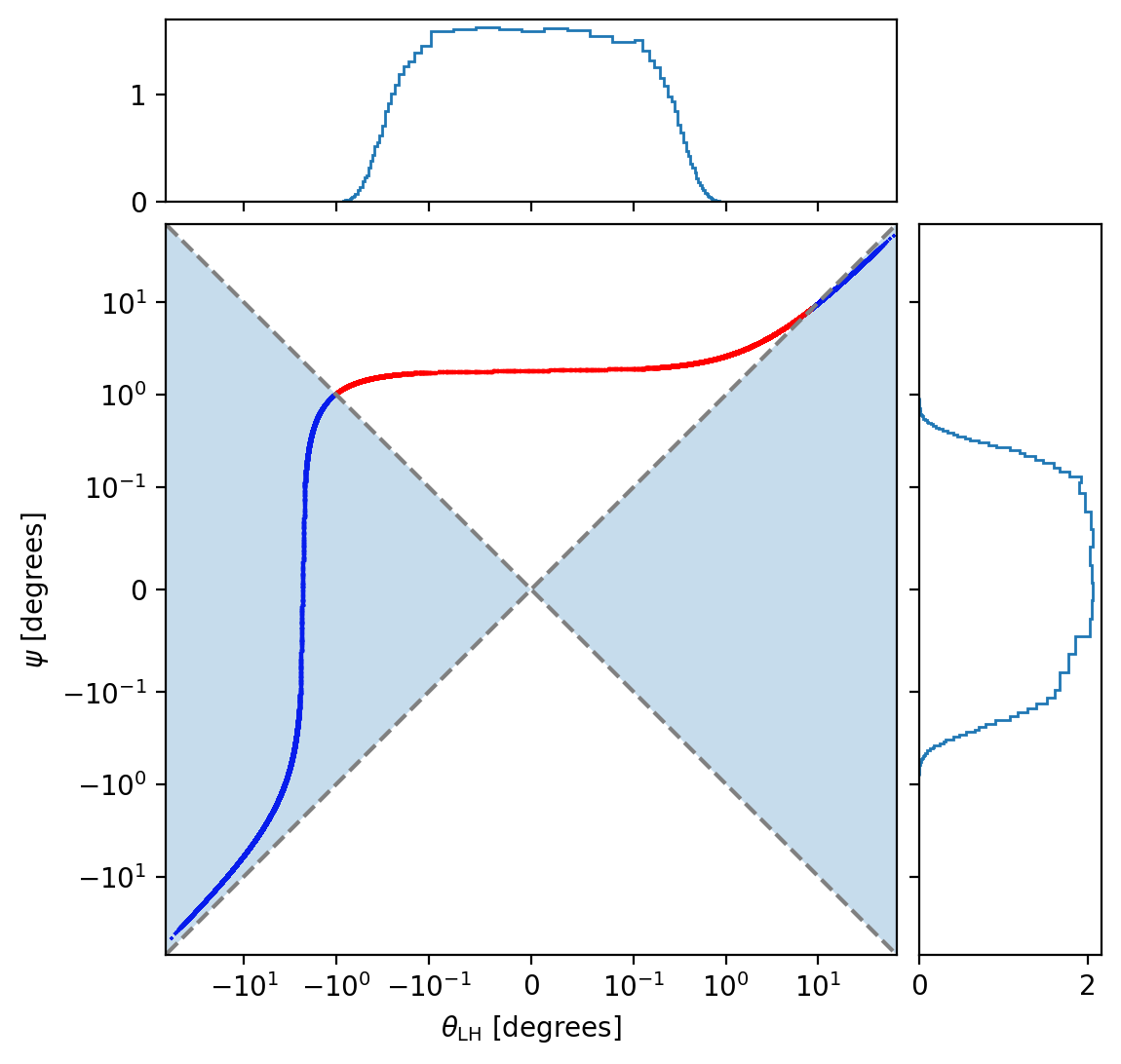}
\includegraphics[width=0.49\textwidth]{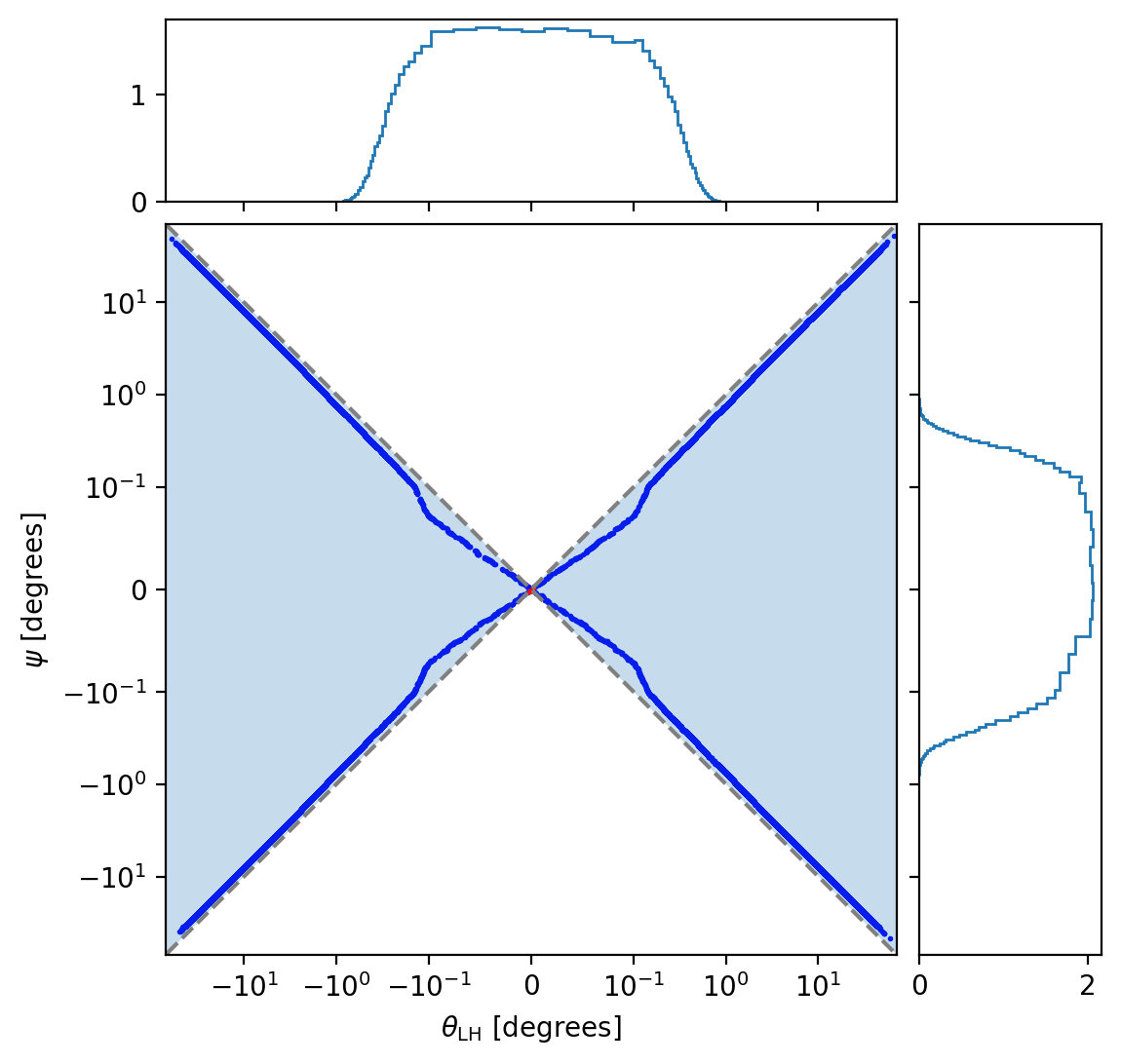}
\caption{Example of randomly drawn $\theta_{\rm LH}$ and $\psi$ angles where a fraction of filaments are nonphysical, violating eq.~\eqref{eq:module_condition}. Left, $\psi$ angles are drawn with a normal distribution with $\mu=1.8^{\circ}$ and $\sigma=11^{\circ}$. Blue points are physical, while red points are nonphysical. Right, the same points after being sorted by their absolute value and re-paired. Now the nonphysical red points are minimal. \label{fig:inverse_transform_sampling}}
\end{figure*}

Nonetheless, there are some caveats with this approach:
\begin{itemize}
    \item In our case, we decided to try continuous probability distributions for $X$. In general, getting an analytical CDF that one can invert is impossible for most common continuous distributions since it involves the integration of the probability density function (PDF) of $X$. As an alternative, we can use the percent point function implemented in the \textsc{scipy.stats} module~\cite{2020NatMe..17..261V} to approximate the inverse of the CDF with percentiles for all the common distributions.
    
    \item While $\theta_{\rm LH} \in \mathcal{N}(\mu=0, \sigma^2=\text{rms}(\theta_{\rm LH})^2)$, for some choices of the $\psi$ random angle where there is too much asymmetry, some of the randomly generated $(\theta_{\rm LH}, \psi)$ pairs will violate our condition eq.~\eqref{eq:module_condition}. Fig.~\ref{fig:inverse_transform_sampling} illustrates this when $\psi \in \mathcal{N}(\mu=1.8^{\circ}, \sigma^2=(11^{\circ})^2)$ and $\theta_{\rm LH} \in \mathcal{N}(\mu=0, \sigma^2=(14^{\circ})^2)$. The left-side panel shows $10^5$ random pairs $(\theta_{\rm LH}, \psi)$ generated with the inverse transform sampling. The light blue area shows the allowed space where condition eq.~\eqref{eq:module_condition} is true. The blue points show the pairs that fulfill the condition, while the red points show the pairs that violate it. In the smaller panels at the top and right side, we can see the histogram of the $\theta_{\rm LH}$ and $\psi$ angles, respectively. A simple way to reduce the number of random pairs that violate the condition is to sort $\theta_{\rm LH}$ and $\psi$ separately by their absolute value, and then re-pair each preserving this order. This is shown in the right-side panel of Fig.~\ref{fig:inverse_transform_sampling}, where most of the random pairs are blue, fulfilling the condition of eq.~\eqref{eq:module_condition}, and only a very small fraction of points still violate it. For this example, $\sim 25$ percent of the random pairs violate the condition initially, while only 0.06 percent of pairs do after the sorting procedure. Note that the PDFs of both $\theta_{\rm LH}$ and $\psi$ are preserved.
    
    \item Depending on how asymmetric the $\psi$ distribution is, after the procedure described above, a small fraction of $(\theta_{\rm LH}, \psi)$ pairs will still violate eq.~\eqref{eq:module_condition}. We perform rejection sampling by creating a new batch of random pairs, sorting them by absolute value, and using them to replace the bad pairs in the original batch of pairs, until all of the $(\theta_{\rm LH}, \psi)$ pairs fulfill the condition. For very asymmetric $\psi$ distributions, these geometric limitations will persist, and the distribution of $\psi$ angles will have ranges of values that are impossible to produce given the geometric constraints. Fig.~\ref{fig:psi_distribution} shows in orange an example histogram for a 10M-filament population where the injected asymmetry makes some $\psi$ ranges unphysical (eq.~\ref{eq:module_condition}), producing holes in the distribution, e.g. the cutoff at $\psi \sim 40^{\circ}$ .
\end{itemize}

For every filament, we perform two consecutive rotations, first by an angle $\theta_{\rm LH}$ and then by an angle $\phi$. However, in this case, where we are injecting an asymmetry, we know the $\theta_{\rm LH}$ angle but not what $\phi$ angle is needed to make the filament-magnetic field projected angle have a value of $\psi$. What we do is define an auxiliary function $f(\phi, \theta_{\rm LH}, \psi)$, which rotates the filament by $\theta_{\rm LH}$, then by $\phi$, and calculates a projected angle $\psi'$. Finally, it returns $\psi'-\psi$, the difference between our target angle and the internally-calculated projected angle. Hence, we want to know for fixed $\theta_{\rm LH}$ and $\psi$, at which $\phi$ our function $f$ is zero. In other words, we want to know the roots of the function $f(\phi_i)=0$. There are two roots, i.e. two values of the azimuthal rotation $\phi_i$ that will give identical $\psi$. As seen from an observer, this would be a near-side and a far-side angle. We use the numerical root finding tools from \textsc{gsl}~\cite{galassi2018scientific} to find the two $\phi_i$ angles, and for every filament, we choose one of the two at random.

The probability distribution used to draw $\psi$ can be anything in theory, but for many possible distributions, the angles will be incompatible with each other given the restrictions of the filament geometry, such as the constraint described by eq.~\eqref{eq:module_condition}. In this work, we test two distributions: the Asymmetric Laplace distribution (ALD) and an off-center normal distribution.

\subsubsection{Asymmetric Laplace distribution} \label{sec:ald}

\begin{figure}
    \includegraphics[width=1.0\columnwidth]{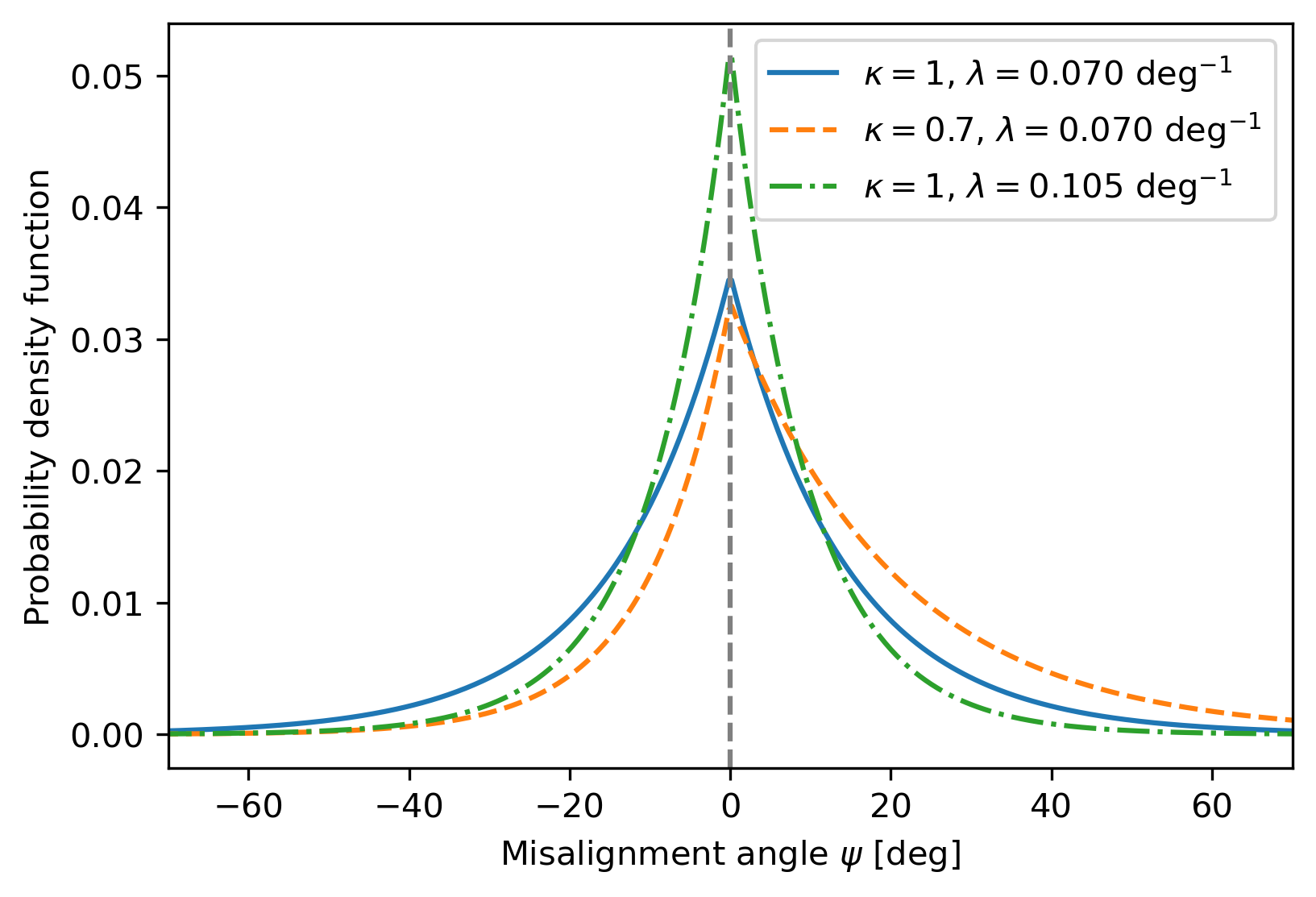}
    \caption{Example PDFs for the Asymmetric Laplace distribution centered at zero. $\kappa=1$ generates equal amounts on the positive and negative sides. $\kappa<1$ generates an excess of positive values. The width of the distribution is proportional to $1/\lambda$.
    \label{fig:ald_examples}
    }
\end{figure}

This probability distribution is defined by the PDF~\cite{ALD}
\begin{equation}
    f(x;\mu,\lambda,\kappa) = \frac{\lambda}{\kappa+1/\kappa} 
    \begin{cases} 
        \exp((\lambda/\kappa)(x-\mu)) & \text{if $x<\mu$} \\
        \exp(-\lambda \kappa (x-\mu) ) & \text{if $x \geq \mu$}
    \end{cases} \text{,}
\end{equation}
where $\mu$ controls the location, $\kappa$ the asymmetry, and $\lambda$ the scale. The width of the ALD is proportional to $\lambda^{-1}$. If the reader is unfamiliar with the ALD, in Fig.~\ref{fig:ald_examples} we show examples of the PDF for various parameters.

In our study, we set $\mu=0$ and vary the $\psi$ random variable with $\kappa$ and $\lambda$, therefore introducing asymmetry by skewing the distribution rather than by shifting the mean. $\kappa=1$ represents a 50/50 split between positive and negative $\psi$, while $\kappa=0.816$ represent an approximately 60/40 split. Fig.~\ref{fig:psi_distribution} shows in orange an example histogram for a 10M-filament population drawn from an ALD with $\kappa = 0.883$, $\lambda=0.1396$\,deg$^{-1}$.

\subsubsection{Normal distribution} \label{sec:normal}

Another option to create asymmetry is to shift the location of a normal distribution slightly towards positive values so that there will be an asymmetry of positive versus negative $\psi$ angles. In this case, the two parameters are $\mu$ for the location and $\sigma$ for the scale of the distribution. Fig.~\ref{fig:psi_distribution} shows in green an example histogram for a 10M-filament population drawn from a normal distribution with $\mu = \dotdeg{1.778}$ and $\sigma=12^{\circ}$.

\section{$\hat{\psi}$ estimators} \label{sec:psi_estimators}

\citetalias{2023ApJ...946..106C} defines map-based and cross-spectra estimators for measuring the misalignment angle $\psi$ that rely on constructing a dust template from HI observations. The HI-derived dust template is built by measuring the linear structures with the methods mentioned in Sec.~\ref{sec:filament_measuring}, assuming a perfect filament-magnetic field alignment, obtaining the HI-measured dust polarization angle, and integrating along the frequency spectrum in velocity bins~\cite{2019ApJ...887..136C,2021ApJ...919...53C}. Complementarily, the millimeter observations by \textit{Planck} in the 353\,GHz channel measure the dust polarization angle directly, with the difference between the two angles quantifying the magnetic misalignment.

\subsection{Map-based estimator} 

A map-based estimator for $\psi$ using a region of the sky with multiple pixels is defined in~\citetalias{2023ApJ...946..106C}, based on a modification to the projected Rayleigh statistic~\cite{2018MNRAS.474.1018J}. The estimator is
\begin{equation} \label{eq:map_based_estimator}
    \hat{\psi_0} = \frac{1}{2} \atantwo(B,A)
\end{equation}
with 
\begin{eqnarray}
    A \equiv \frac{1}{W} \sum_{\hat{\vect{n}}} w(\hat{\vect{n}}) (c_{\rm HI} c_{\rm d} + s_{\rm HI}s_{\rm d}) \\
    B \equiv \frac{1}{W} \sum_{\hat{\vect{n}}} w(\hat{\vect{n}}) (c_{\rm HI} s_{\rm d} - s_{\rm HI}c_{\rm d}) \text{,}
\end{eqnarray}
where $c_x \equiv Q_x/P_x$, $s_x \equiv U_x/P_x$, $P_x=\sqrt{Q_x^2+U_x^2}$, $x \in [{\rm HI}, {\rm d}]$. 

When estimating eq.~\eqref{eq:map_based_estimator}, we need an estimate of the signal-to-noise ratio (SNR) to use as weights $w(\hat{\vect{n}})$. We use the covariance per pixel from the \textit{Planck} 353\,GHz channel map. Following the same weighting scheme from~\citetalias{2023ApJ...946..106C}, the polarization covariance per pixel is given by 
\begin{equation}
    \Delta P_{353}^2 = \frac{{\rm cov}(Q,Q) Q^2 + {\rm cov}(U,U) U^2 + 2 {\rm cov}(Q,U) Q U }{\sqrt{Q^2+U^2}} \text{,}
\end{equation}
where ${\rm cov}(X,Y)$ represents the covariance between the fields $X$ and $Y$. The noise in the HI4PI is assumed to be homogeneous, so the SNR is proportional to the signal. Therefore, the total weight is the multiplication of both SNR estimates,
\begin{equation}
    w(\hat{\vect{n}}) = \frac{P_{\rm d}(\hat{\vect{n}})}{\Delta P_{353}(\hat{\vect{n}})} P_{\rm HI}(\hat{\vect{n}}) \text{.}
\end{equation}
\citetalias{2023ApJ...946..106C} did not present this prescription due to conciseness. In the case of our filament model, which is pure signal, we simply use $w(\hat{\vect{n}}) = P_{\rm d}(\hat{\vect{n}}) P_{\rm HI}(\hat{\vect{n}})$ as our weight.

\subsection{Cross-spectra estimators} \label{sec:psi_estimators_crosscl}

We can define cross-spectra-based estimators for large sky fractions. These four estimators for the scale-dependent misalignment angle $\psi_{\ell}$ are given by
\begin{equation} \label{eq:psi_spectra_estimators}
\begin{split}
    \tan(2 \hat{\psi}_{1,\ell}) =& \phantom{-}\frac{C_{\ell}^{E_{\rm HI}B_{\rm d}}}{C_{\ell}^{E_{\rm HI}E_{\rm d}}}, \\
    \tan(2 \hat{\psi}_{2,\ell}) = & -\frac{C_{\ell}^{B_{\rm HI}E_{\rm d}}}{C_{\ell}^{B_{\rm HI}B_{\rm d}}}, \\
    \tan(2 \hat{\psi}_{3,\ell}) = & \phantom{-}\frac{C_{\ell}^{T_{\rm HI}B_{\rm d}}}{C_{\ell}^{T_{\rm HI}E_{\rm d}}}, \\
    \tan(2 \hat{\psi}_{4,\ell}) = &\phantom{-}\frac{C_{\ell}^{T_{\rm d}B_{\rm d}}}{C_{\ell}^{T_{\rm d}E_{\rm d}}} \text{.}
\end{split}
\end{equation}

Analogous to eq.~\eqref{eq:EB_from_dust}, $\hat{\psi}_{1,\ell}$ shows a ratio between $EB$ and $EE$ spectra. However, both ratios are different because they involve different spectra, $E_{\rm d} B_{\rm d}/E_{\rm d} E_{\rm d}$ versus $E_{\rm HI} B_{\rm d}/E_{\rm HI} E_{\rm d}$. Furthermore, eq.~\eqref{eq:EB_from_dust} is suppressed by the factor $A_{\ell} \ll 1$, while in eq.~\eqref{eq:psi_spectra_estimators} the HI-derived fields have no such effect. On the other hand, $\hat{\psi}_{4,\ell}$ is the same as eq.~\eqref{eq:psi_TB_TE}. 

Three out of four of these estimators are cross-correlations between the dust and HI-derived template, which have independent noise realizations. For estimation on real data, we use the full \textsc{commander} dust maps. $\hat{\psi}_{4,\ell}$ depends only on dust maps, and therefore we cross-correlate the two half-mission dust maps from \textsc{commander} to avoid a potential noise bias. Also, this estimator has contributions from the non-filamentary structure in dust and therefore might introduce some systematics that do not reflect the filament misalignment angle.

\subsection{Construction of HI-based dust template} \label{sec:filament_measuring}

As described in Refs.~\cite{2021ApJ...919...53C} and~\citetalias{2023ApJ...946..106C}, the angle $\psi$ is measured from both direct observation of dust and a $TQU$ template constructed from measuring filaments from HI data. In this section, we summarize two methods for achieving this: the Rolling Hough Transform (RHT)~\cite{2014ApJ...789...82C, 2020ascl.soft03005C} and the Hessian method~\citepalias{2023ApJ...946..106C}. Then, we describe how we do this calculation for our simulated filament population.

\subsubsection{Rolling Hough Transform} \label{sec:rht}
The RHT is a machine vision algorithm that measures the orientation of linear structures in a 2D image. In particular, applied to HI data, the RHT measures the intensity of HI structure as a function of orientation. Since HI spectroscopic data contains information on position $\hat{\vect{n}}$ and velocity bin $v$ due to Doppler shift, if we run it on every spectral channel map, we will have information on the intensity of structure as a function of position, velocity and angle. We refer the reader to Refs.~\cite{2019ApJ...887..136C,2024ApJ...961...29H} for specifics.

To produce an HI-derived $TQU$ template, at each velocity bin we measure linear structures at a limited number of orientations around the circle that are longer than some scale. Then, we sum across overall velocities of interest to obtain a $TQU$ template. Ref.~\cite{2024ApJ...961...29H} extended the RHT to work directly on the surface of the sphere using convolution on harmonic space, and that is the implementation we use in this work.\footnote{\url{https://github.com/georgehalal/sphericalrht}}

\subsubsection{Hessian method} \label{sec:hessian}

The HI-derived dust template can be built through an alternative method that exploits the local Hessian matrix, which contains information about the second derivative and curvature. Filaments are identified through areas of negative curvature. We refer the reader to Refs.~\cite{2023ApJ...946..106C,2024ApJ...961...29H} for details on the method. It is applied to a velocity bin and integrated along the line of sight to obtain a $TQU$ template.

\subsubsection{Filament model proxy for HI spectral observations} Refs.~\cite{2021ApJ...919...53C} and~\citetalias{2023ApJ...946..106C} used real data, \textit{Planck} mm dust and HI spectral observations from the HI4PI survey. For our dust filament model, we produce a dust map, but we obviously do not have something similar to HI spectral observations. Instead, we follow a procedure to obtain an analogous third dimension along the line of sight. While not having Doppler shift velocity, we can make intensity maps of our cubic volume filament population in concentric shells at equidistant radii. 

When running our filament model to create a population and a simulated $TQU$ map, we also save the $T$ field in one of 20 radial bins depending on the radial distance from the observer to filament, between 0 and 160\,pc. This way, we also obtain 20 $T$ maps of the concentric shells for the same simulated filament population. Then, we run either the Hessian or RHT method over these 20 maps. This morphology-derived template will be labeled ``HI'' throughout this work in analogy to the HI derivation done with real HI observations in Refs.~\cite{2019ApJ...887..136C,2021ApJ...919...53C,2023ApJ...946..106C}.% A caveat to keep in mind is that our radial shells are not truly independent from the dust map, while the real HI data is truly an independent probe from real dust observations.

\begin{figure*}
    \includegraphics[width=1.0\textwidth]{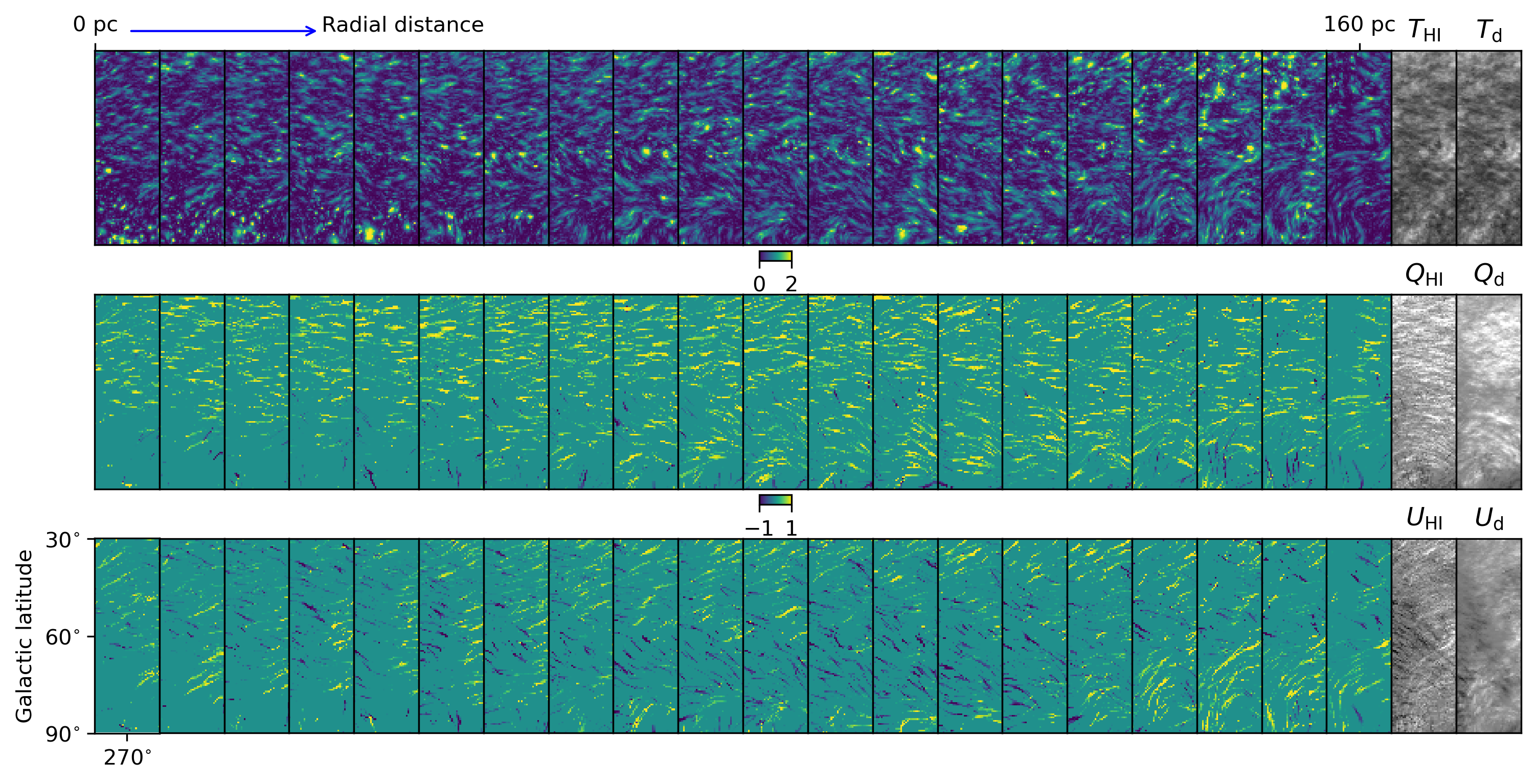}
    \caption{How we construct our morphology-derived dust template for a filament model map. Each panel is a rectangular cut centered in Galactic coordinates $(l,b)=(270^{\circ},+60^{\circ})$ with size $20^{\circ} \times 60^{\circ}$. The first row is $T$ calculated with the filament model in each concentric radial shell. The second and third row are the $Q$ and $U$ fields derived using the RHT estimation. The color-scale maps show maps individually for the 20 radial shells, increasing in radius towards the right. The gray-scale maps show the sum of the 20 radial shells, labeled HI, compared to the directly simulated dust map, labeled d. The units are arbitrary.
    \label{fig:HI_template}
    }
\end{figure*}

Fig.~\ref{fig:HI_template} shows this procedure for a filament model realization. Each row corresponds to the $T$, $Q$, and $U$ fields, while the color-scale maps show the 20 radial shells from 0 to 160\,pc increasing towards the right. The sum across the radial shells is shown in grayscale to the right, next to the dust realization. As can be seen, the $T$ fields are equivalent by definition, while the $Q/U$ fields are very correlated depending on how good the Hessian method/RHT approximation of filament orientation is.\footnote{For Fig.~\ref{fig:HI_template}, the pixel-pixel correlation coefficient between $Q_{\rm HI}$ and $Q_{\rm d}$, degraded to $N_{\rm side}=256$, is 0.822. For $U$, the same coefficient is 0.818.} This illustrates the limitations of filament-finding methods since our simulated signal is entirely made up of filaments and is perfectly known, yet the correlation is not perfect due to limiting factors in the methods, such as pixelization.
\section{Best-fit model results} \label{sec:results}
In this section, we detail the result from calibrating our model to the $\psi$ angle measurements with \textit{Planck} and HI data. We describe our fitting procedure, we show the observations we will be fitting to, and then detail our results. Finally, we present our fiducial dust model with non-zero parity-violating spectra.

\subsection{Fitting the model} \label{sec:fitting_a_model}

To determine which distribution of $\psi$ angles fits the sky observations, we calculate the $\psi$ estimators described in Sec.~\ref{sec:psi_estimators} for a simulated filament model realization, and we compare that to the estimators calculated over the real observations, as shown in~\citetalias{2023ApJ...946..106C}. 

To estimate the uncertainty of the real observations, we use 50 simulations from that same paper. Sec.~5 of~\citetalias{2023ApJ...946..106C} described how these mock skies are produced. The simulations include realizations of Gaussian noise and dust, and a constant-across-realizations filamentary HI component, derived from the HI4PI data using the Hessian method. The simulations are built to explicitly replicate the two-point correlations of the real sky, i.e., the angular cross-power spectra between combinations of dust and the HI-derived template of the simulations are the same as the one calculated from the true sky. A simulated map $S$ consists of a Gaussian noise realization matching the \textit{Planck} 353\,GHz channel sensitivity, a Gaussian dust that matches the power spectra of dust calculated from \textsc{commander} as described in Sec.~\ref{sec:planck_data}, and the HI component, which is modulated in harmonic space by an ad-hoc $\ell$-dependent transfer function such that $\mathcal{D}^{X_{\rm HI} X_{\rm S}}_{\ell} = \mathcal{D}^{X_{\rm HI} X_{\rm d}}_{\ell}$. The Gaussian dust and noise templates are isotropic initially. The \textsc{commander} dust template smoothed to $\dotdeg{14.7}$ is used to spatially modulate the Gaussian dust with the goal of mimicking the anisotropy of the real Galactic dust, by estimating the unbiased dust spectra from half-mission maps on large patches of the sky corresponding to the pixels of a \textsc{healpix}\footnote{\url{https://healpix.sourceforge.io/}}~\cite{2005ApJ...622..759G} map with $N_{\rm side}=8$. The Gaussian noise is also modulated spatially in a similar way by estimating the noise bias spectra through the subtraction of auto spectra and the unbiased dust spectra. All of the maps are masked with the 70\% Galactic plane mask before transforming to harmonic space, and as such the resulting mock skies are well-defined inside the mask only.

We fit our model using observations and we construct a likelihood that adjusts all the observables jointly, namely the five estimators defined in Sec.~\ref{sec:psi_estimators}. %We assume that the data is Gaussian distributed. We form a likelihood given by
We assume the following Gaussian likelihood
\begin{equation} \label{eq:likelihood}
    2 \ln \mathcal{L}(\vect{d} | \vect{p}) = -\chi^2 = -\vect{d} \matr{C}^{-1} \vect{d} \text{,}
\end{equation}
where $\vect{d}$ is our data residual vector, with the form
\begin{equation}
\vect{d} = \left[ \Delta \psi_{i=0}^{b=0},..., \Delta \psi_{i=0}^{b=N_{b}-1}, ..., \Delta \psi_{i=N_{\rm e}-1}^{b=0},..., \Delta \psi_{i=N_{\rm e}-1}^{b=N_{b}-1} \right] \text{,}
\end{equation}
where $N_{b} = 5$ is the number of multipole bins in the range $\ell=200-700$\footnote{We choose the lower end of this range to be $\ell=200$ since at these scales and smaller the filament model polarization looks like a consistent power law that can be directly compared to the \textit{Planck}-measured dust.} with width $\Delta \ell = 100$, $N_{\rm e}$ is the number of $\psi$ estimators used. Therefore, $\vect{d}$ is a vector with length $N_{\rm e}N_b$. $\Delta \psi_{i}^{b} = \psi_{i,b}^{\rm observations} - \psi_{i,b}^{\rm model}(\vect{p})$, i.e., the difference between estimator $i$ at bin $b$ from real sky observations (as estimated in~\citetalias{2023ApJ...946..106C}) and the same from a filament dust template with parameters $\vect{p}$.\footnote{While $\hat{\psi}_{1..4}$ are simply ratios of angular power spectra, to build the $\hat{\psi}_{0}$ estimator we filter the maps with a bandpass harmonic filter with $\Delta \ell=100$, apply eq.~\eqref{eq:map_based_estimator} summing over pixels within the mask, and that gives us the data point for the corresponding $\ell$ range. We then repeat for every range between $\ell=200$ to $700$.} In this case, $\vect{p}$ are the parameters of the probability distribution for the angle $\psi$. We adopt uniform priors for the parameters $\vect{p}$ in all cases since we do not have a well-motivated physical expectation for the distribution of the $\psi$ angles. 

The $C_{\ell}$ band powers are very close to Gaussian since we are working at high $\ell > 200$. We checked that the spectra-based estimators defined in eq.~\eqref{eq:psi_spectra_estimators} are reasonably close to Gaussian. For the map-based estimator $\hat{\psi}_0$ of eq.~\eqref{eq:map_based_estimator}, pure noise is distributed uniformly, but our null hypothesis is defined by the \emph{combination} of perfect alignment together with noise fluctuations. In that case, the distribution is localized near the aligned angle. For simplicity, we assume Gaussianity for this estimator. Given these limitations, we abstain from making strong claims on inferring the suitability of specific models versus others.

The covariance matrix $\matr{C}$ is estimated empirically from 50 realizations of the mock skies described above. While simulations are produced applying spatial modulation to Gaussian isotropic fields, we expect the mode coupling to be relevant only on the largest scales, well below our lower limit of $\ell=200$, given that the modulating template is smoothed to a scale of $\dotdeg{14.7}$. Then, we expect independent modes in the multipole range of interest, and therefore we null the covariance between different multipole bins to avoid spurious correlations while allowing covariance across different estimators. While the 50 mock skies from \citetalias{2023ApJ...946..106C} lack the realism from non-Gaussian structures present in dust and therefore should not be used for making strong claims of statistical inference, we believe they represent a good approximation of the covariance, which we need to weigh our observables.

\subsection{$\hat{\psi}$ estimators from true sky observations} \label{sec:results_observations}

\begin{figure}
    \includegraphics[width=1.0\columnwidth]{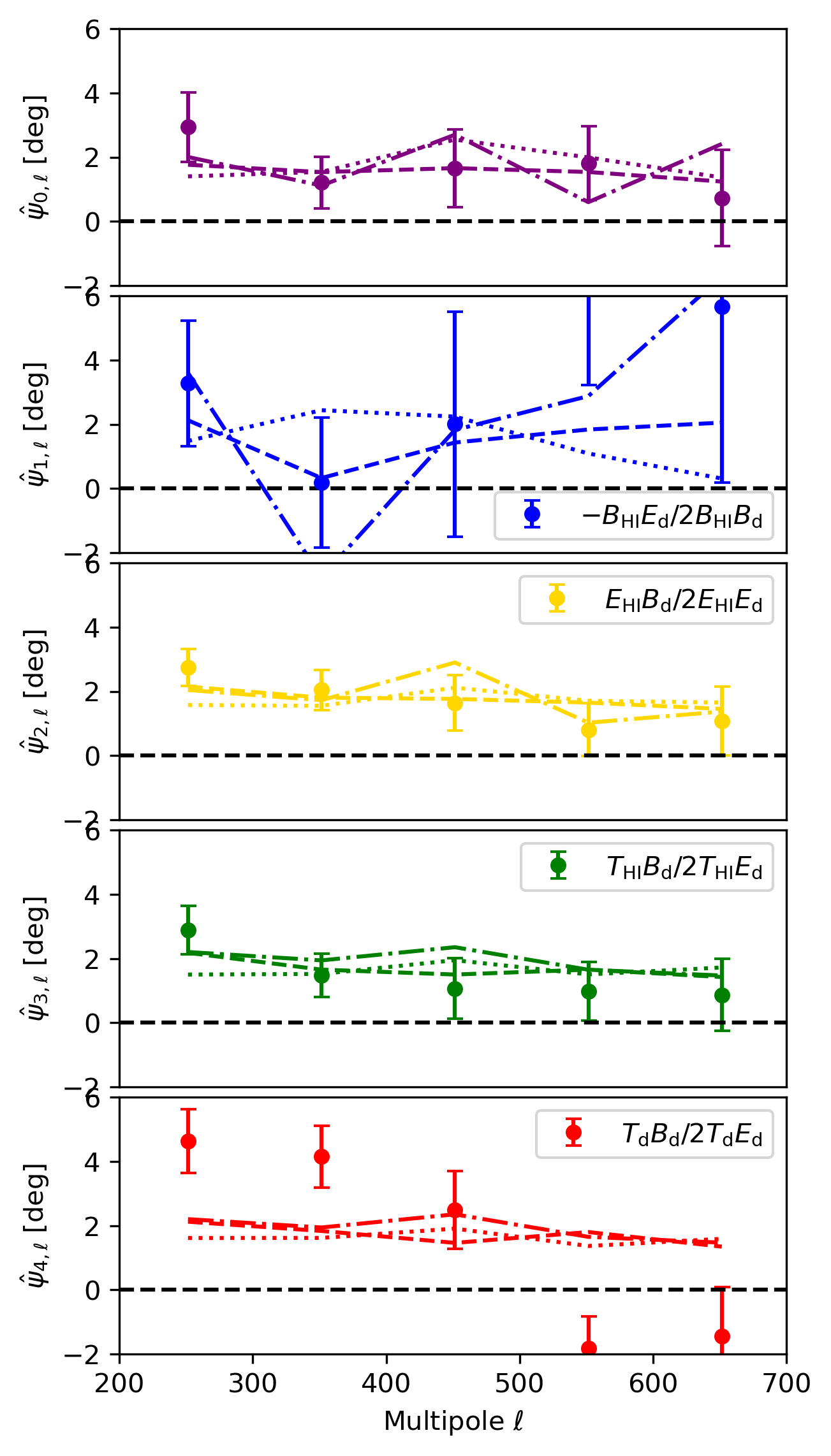}
    \caption{The five $\psi$ estimators described in Sec.~\ref{sec:psi_estimators}. The data points are the estimators for real observations (see~\citetalias{2023ApJ...946..106C}). The dashed line corresponds to the best-fit fiducial model (Sec.~\ref{sec:fiducial_bestfit}), while the dotted and dash-dotted lines correspond to the best-fit alternative models using the normal distribution for generating $\psi$, with the Hessian and RHT methods to construct the HI-derived template, respectively. The estimators for our filament model are a single realization, and therefore subjected to cosmic variance.
    \label{fig:psi_estimators}
    }
\end{figure}

\citetalias{2023ApJ...946..106C} finds that the angle $\psi$ has a value of $\sim 2^{\circ}$ in the multipole range $100 \leq \ell \leq 700$, being roughly scale-independent. Fig.~\ref{fig:psi_estimators} shows in filled circles the five $\psi$ estimators described in Sec.~\ref{sec:psi_estimators} and shown in~\citetalias{2023ApJ...946..106C}. These are calculated with the \textit{Planck} \textsc{commander} dust template, a HI-derived template using the Hessian method, using the 70\% Galactic mask. The error bars are calculated from the standard deviation across 50 realizations of the mock skies described in Sec.~\ref{sec:fitting_a_model}.

\subsection{Calibrating the filament model} \label{sec:results_models}

First, we will describe a fiducial dust filament model that uses the ALD (described in Sec.~\ref{sec:ald}) to generate the random $\psi$ angles. Then, we will change different aspects of the methodology to test how robust our modeling is.

\subsubsection{Fiducial model} \label{sec:fiducial_bestfit}

\begin{table*}
    \caption{Different parameters for the filament model used in this work as compared to Table~1 of~\citetalias{2022ApJ...928...65H}. \label{table:filaments_parameters}}
    \begin{ruledtabular}
    \begin{tabular}{l l l}
        Parameter & Symbol & Value \\
        \hline
        Number of filaments for 70\% Galactic mask & $N_{\rm fil}$ & 50 million \\
        Filament density & $n_{\rm fil}$ & 5583\,deg$^{-2}\times [I_{\rm dust}/({\rm MJy\ sr}^{-1})]$ \\
        Filament length, Pareto distribution & $p(L_a)$ & $\propto L_a^{-2.542}$  \\
        Filament axis ratio & $\epsilon$ & $0.137 (L_a/L_a^{\rm min})^{+0.165}$   \\
        Filament misalignment angle dispersion & rms($\theta_{LH}$) & $14^{\circ}$  \\
        Polarization fraction geometric dependence & $f_{\rm pol}$ & $\propto (L_a/L_a^{\rm min})^{-0.102}$  \\
    \end{tabular}
    \end{ruledtabular}
\end{table*}

The starting point for our fiducial model is the setup presented in~\citetalias{2022ApJ...928...65H} and summarized in Sec.~\ref{sec:summary}. That model used $\text{rms}(\theta_{\rm LH}) = 10^{\circ}$, while in this work we use $\text{rms}(\theta_{\rm LH}) = 14^{\circ}$. The reason for this is to allow a wider range of values for the $\psi$ angle since the latter is physically restricted by the value of $\theta_{\rm LH}$, as explained in Section~\ref{sec:asymmetric_psi}. There is a degeneracy between $\text{rms}(\theta_{\rm LH})$ and $\epsilon$ (the aspect ratio of a filament) parameters, since thinner filaments can produce similar spectra if less aligned, as noted in~\citetalias{2022ApJ...928...65H}. This change means that other parameters of the filament model also must change. Furthermore, we calibrate the dust angular power spectra to the Galactic 70\% mask, while~\citetalias{2022ApJ...928...65H} calibrated with respect to the LR71 \textit{Planck} mask. Table~\ref{table:filaments_parameters} summarizes some of the parameters used to generate the filament model. This table only shows the values that are different with respect to Table~1 of~\citetalias{2022ApJ...928...65H}. Note that we use 50 million filaments that are placed according to the 70\% Galactic mask rather than simulating the full sky. We do this to save computing time by not generating tens of millions of filaments inside the Galactic plane that will be masked anyway and not used.

\begin{figure}
    \includegraphics[width=1.0\columnwidth]{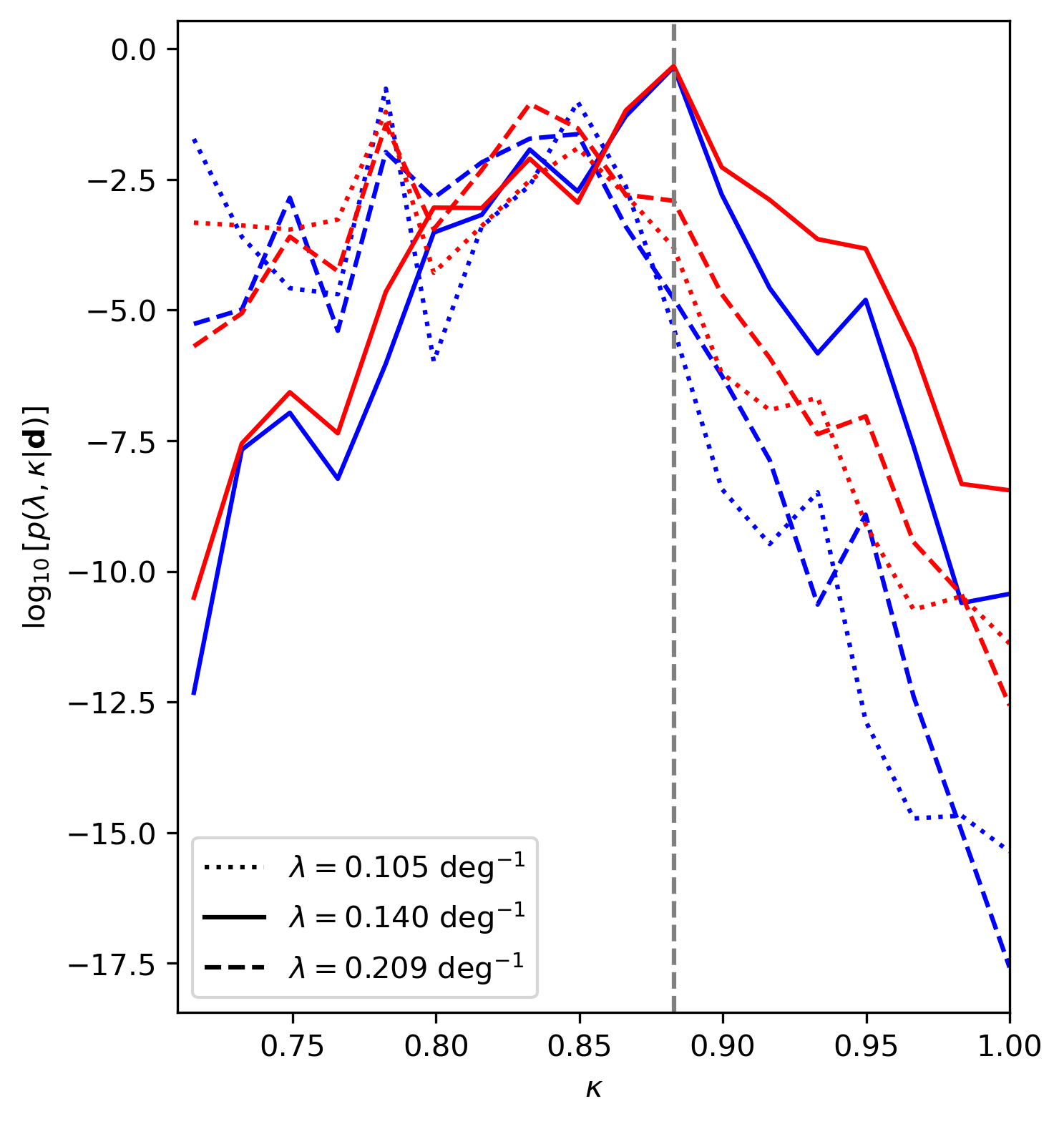}
    \caption{Posterior probability $p(\lambda,\kappa | \vect{d})$ from fitting our filament model to sky observations using $\psi$ estimators, shown as a function of $\kappa$ for a few fixed values of $\lambda$. This is for our fiducial model, which uses the ALD for generating $\psi$ angles and the Hessian method to reconstruct filaments. The blue curves use all five estimators defined in Sec.~\ref{sec:psi_estimators}, while the red curves exclude the last estimator from eq.~\eqref{eq:psi_spectra_estimators}. The best-fit model is marked with a vertical dashed line.
    \label{fig:poesterior_fiducial}
    }
\end{figure}

Generating a dust filament realization is relatively expensive, taking a few hours in a node with $\sim 50$ cores. Hence, we cannot freely sample and maximize the likelihood with, e.g., Markov chain Monte Carlo (MCMC) methods. Instead, we use a predefined 2D grid of the 2 parameters of the ALD, the asymmetry $\kappa$ and the scale $\lambda$. We run a wide range for both parameters, to give us a general idea of how good or bad a model for $\psi$ will fit the sky observations. We run realizations of the filament model at $N_{\rm side}=512$ for 10 values of $\lambda$ in the range $0.087-0.244$\,deg$^{-1}$, and 18 values for $\kappa$ in the range $0.716-1.0$, giving us a grid of 180 $(\kappa,\lambda)$ parameter values. To construct the HI-derived dust template, we use the Hessian method. Our model for each combination of parameters is actually a single realization of the model, so our fitting is subjected to cosmic variance. Ideally, we would run many realizations of the model for the same parameters and average, but this is impractical. The posterior probability $p(\lambda,\kappa | \vect{d})$, derived from eq.~\eqref{eq:likelihood}, as a function of $\kappa$ is shown for a few fixed values of $\lambda$ in Fig.~\ref{fig:poesterior_fiducial}. The blue curves show the fit to data using all five $\hat{\psi}$ estimators defined in Sec.~\ref{sec:psi_estimators}, while the red curves show the case where the $\hat{\psi}_{4,\ell}$ estimator, which only depends on the dust template, is excluded.

We justify this approach on the fact that when applied to observations of the sky, this estimator measures all dust morphology, both filamentary and non-filamentary, while the other estimators are sensitive to filamentary structure by cross-correlating dust with HI. From the bottom panel of Fig.~\ref{fig:psi_estimators}, this estimator somewhat disagrees with the other estimators, showing an angle $\psi \sim 5^{\circ}$ at the largest scales, which subsequently goes negative for scales $\ell \gtrsim 500$. Returning to Fig.~\ref{fig:poesterior_fiducial}, we note that the morphology of the posterior probability in both cases is very similar, but with smaller values which reflect a higher $\chi^2$ for the same combination of parameters, since the pure dust $\hat{\psi}_{4,\ell}$ estimator does not agree with the constant $\psi_{\ell} \sim 2^{\circ}$ that the other estimators seem to measure. Also, we note a relatively weak constraint to the $\lambda$ parameter, while a much stronger constraint to $\kappa$.

The best-fit filament model has $\kappa=0.883$ (equivalent to 56.2\% of $\psi$ angles with positive values) and $\lambda = 0.1396$\,deg$^{-1}$, shown as a vertical dashed line in Fig.~\ref{fig:poesterior_fiducial}. This is for both using all five estimators as well as discarding the pure dust $\hat{\psi}_{4,\ell}$ estimator. The reduced-$\chi^2$ are 1.81 and 0.53, respectively. The five $\hat{\psi}$ estimators for this best-fit model are shown in Fig.~\ref{fig:psi_estimators} as the dashed lines. We note the estimators measured from the best-fit fiducial model agree very well with the estimators measured from the true sky, except in the case of the bottom panel of the figure, for the $\hat{\psi}_{4,\ell}$ estimator. Based on this result, we will use only four $\psi$ estimators and drop the pure dust $\hat{\psi}_{4,\ell}$ estimator.

\subsubsection{Distributions of $\psi$ angles: ALD vs. normal} \label{sec:fit_normal}

\begin{figure}
    \includegraphics[width=1.0\columnwidth]{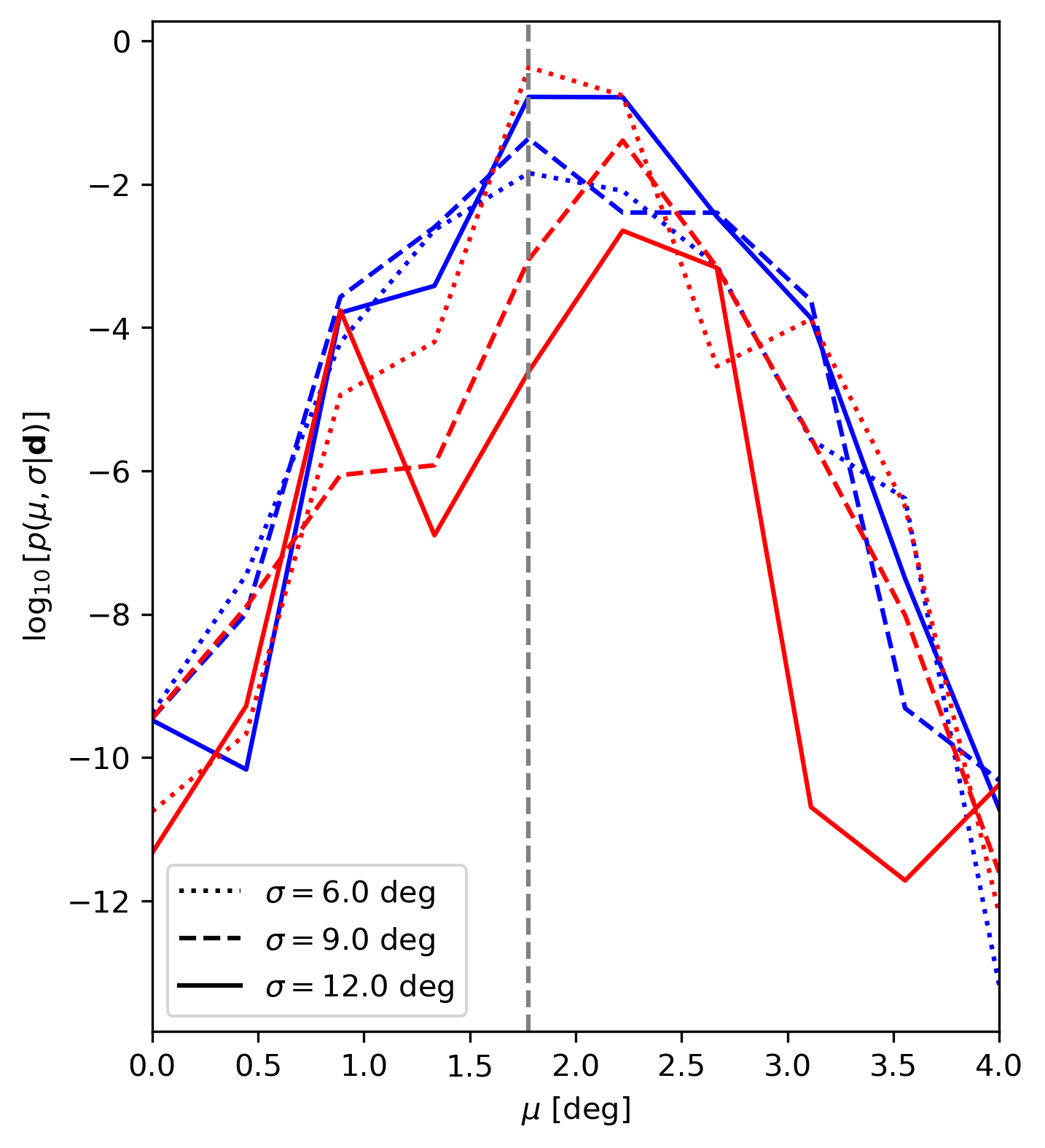}
    \caption{
    Posterior probability for filament models using a normal distribution for the $\psi$ angles, as a function of $\mu$ for a few fixed values of $\sigma$. The blue curves show the use of the Hessian method for the HI-derived template, while the red curves show the use of the RHT. The vertical dashed line marks the best-fit model with $\mu=1.778^{\circ}$ (in the Hessian method case with $\sigma=12^{\circ}$, and in the RHT method case with $\sigma=6^{\circ}$).
    \label{fig:posterior_norm}
    }
\end{figure}

In this section, we switch the ALD for the normal distribution, described in Sec.~\ref{sec:normal}, to generate random $\psi$ angles. The two parameters of the distribution will be the location $\mu$ and the scale $\sigma$. The configuration is the same as for the fiducial model, except we generate a 2D grid of filament realizations for 11 values of $\sigma$ in the range $3-13^{\circ}$, and 10 values of $\mu$ in the range $0-4^{\circ}$, for a total of 110 filament model realizations. The posterior probability for this case, using the Hessian method, is shown as the blue curves in Fig.~\ref{fig:posterior_norm}. The best-fit model has $\mu=\dotdeg{1.778}$ and $\sigma=12^{\circ}$ (equivalent to 55.9\% of positive $\psi$ angles), with a reduced-$\chi^2=1.03$. This is shown as the vertical dashed line in Fig.~\ref{fig:posterior_norm}.

\begin{figure}
    \includegraphics[width=1.0\columnwidth]{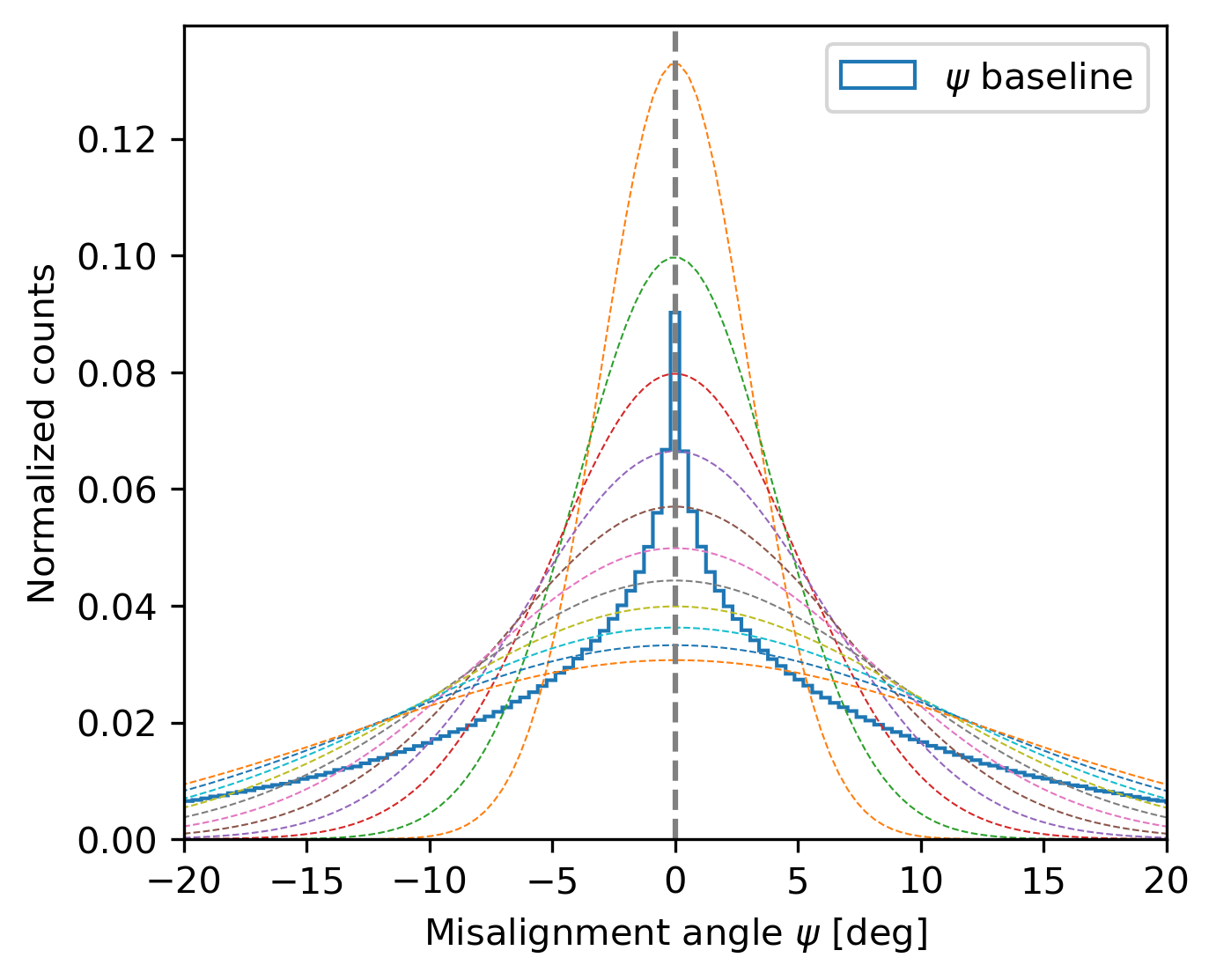}
    \caption{
    Histogram of the $\psi$ angle in the baseline model (in solid blue, the same as shown in Fig.~\ref{fig:psi_distribution}), together with 11 PDFs for the normal distribution with $\mu=0^{\circ}$ and $\sigma=3-13^{\circ}$ (dashed lines).
    \label{fig:psi_distribution_norm}
    }
\end{figure}

The posterior probability seen in Fig.~\ref{fig:posterior_norm} shows a very weak dependency on the $\sigma$ parameter. This reflects the fact that a normal distribution shape does not necessarily match the shape that the filament misalignment angle distribution will naturally take when an asymmetry is not injected artificially. In Fig.~\ref{fig:psi_distribution_norm}, we show the $\psi$ distribution for the filament baseline model, together with the 11 PDFs of the normal distribution used in the model fit, with fixed $\mu=0^{\circ}$ and $\sigma=3-13^{\circ}$. As can be seen from the figure, the exact shape of the normal distribution does not resemble the $\psi$ distribution for any value of $\sigma$. Instead, the amount of magnetic misalignment asymmetry that can be detected with $\psi$ estimators depends mostly on how many filaments have positive versus negative $\psi$ angle, which is controlled mainly by the $\mu$ location. Hence, the fit of our model is mostly $\sigma$-independent.

\subsubsection{HI-based dust template construction: Hessian vs. RHT} \label{sec:hessian_vs_rht}

Another test we can do to check the robustness of our model is to measure filaments with a different method. So far we have shown results using the Hessian method, but we can also use the RHT method.

In this case, for each filament model realization, we run the spherical RHT software from Ref.~\cite{2024ApJ...961...29H} in each of 20 concentric shell $T$ maps, and sum along concentric shells. We use $Z=0.7$, $\theta_{\rm FWHM}=\dotarcmin{40.0}$, and $D_W = \dotarcmin{320.0}$. We use 25 orientations around the circle and $N_{\rm side}=512$. In Appendix~\ref{sec:RHT_study}, we detail why this set of parameters is chosen. Using the ALD for the random $\psi$ angles, and using the same 2D grid of predefined $(\kappa,\lambda)$ parameters from Sec.~\ref{sec:fiducial_bestfit}, we perform a fit of our filament model. The best-fit model has $\kappa=0.783$ and $\lambda=0.1047$\,deg$^{-1}$, with a reduced-$\chi^2=0.60$. We do not show the posterior, but its morphology has the same general shape as the one using the Hessian method (Fig.~\ref{fig:poesterior_fiducial}), although the best-fit model has a slightly more asymmetric distribution of angles, equivalent to 62.0\% of the filaments with positive $\psi$.

We also run the RHT method in the case of using a normal distribution for the random $\psi$ angles. We use the same 2D grid of predefined $(\mu,\sigma)$ parameters from Sec.~\ref{sec:fit_normal}. The result from fitting our model is shown in Fig.~\ref{fig:posterior_norm}, red curves. The best-fit model in this case is for $\mu=\dotdeg{1.778}$ and $\sigma=6^{\circ}$ (equivalent to 61.7\% of positive $\psi$ angles) with a reduced-$\chi^2=1.17$. This is shown as the vertical dashed line in Fig.~\ref{fig:posterior_norm}. We can see in the figure that the morphology of the posterior probability is very similar when comparing using the Hessian versus the RHT method. While the parameters of the best-fit are different in both cases, the posterior in the case of the Hessian method in the position of the RHT method is still close to a local maximum ($p = 0.014$ versus $p = 0.166$ for the global maxima). 

\begin{table*}
    \caption{ 
    Summary of the different best-fit models we found in this paper. 
    \label{table:best_fit_models}
    }
    \begin{ruledtabular}
    \begin{tabular}{l l l l l}
        Prob. distribution for $\psi$ & Parameters & Method for HI-derived dust & Reduced-$\chi^2$ & Percentage of $\psi>0$ \\
        \hline
        Asymmetric Laplace & $\kappa=0.883$, $\lambda=0.1396$\,deg$^{-1}$ & Hessian & $0.53$ & 56.2 \\
        Normal & $\mu=1.778^{\circ}$, $\sigma=12^{\circ}$ & Hessian & $1.03$ & 55.9 \\
        Asymmetric Laplace & $\kappa=0.783$, $\lambda=0.1047$\,deg$^{-1}$ & RHT & 0.60 & 62.0 \\
        Normal & $\mu=1.778^{\circ}$, $\sigma=6^{\circ}$ & RHT & 1.17 & 61.7 \\
    \end{tabular}
    \end{ruledtabular}
\end{table*}

All of the best-fit parameters for the different combinations we try in this section are summarized in Table~\ref{table:best_fit_models}. In all cases, since the filament model being fitted is actually a single realization of a model and therefore still subjected to sample/cosmic variance, it is possible that fluctuations in a particular realization will change the $\chi^2$ of a particular model. As such, the differences in, e.g., the asymmetry in the $\psi$ angle evidenced in Table~\ref{table:best_fit_models}, are not statistically significant. However, what is significant is the preferred asymmetry towards positive $\psi_\ell$ across all models.

\subsection{Best-fit fiducial model: spectra and $EB$ prediction} \label{sec:prediction_for_EB}

Having tested our filament model with different configurations and assessed its robustness, we adopt the fiducial model described in Sec.~\ref{sec:fiducial_bestfit} as our model for the Galactic thermal dust emission with parity-violating statistics calibrated to the sky observations, summarized in the first row of Table~\ref{table:best_fit_models}.

\begin{figure}
    \includegraphics[width=1.0\columnwidth]{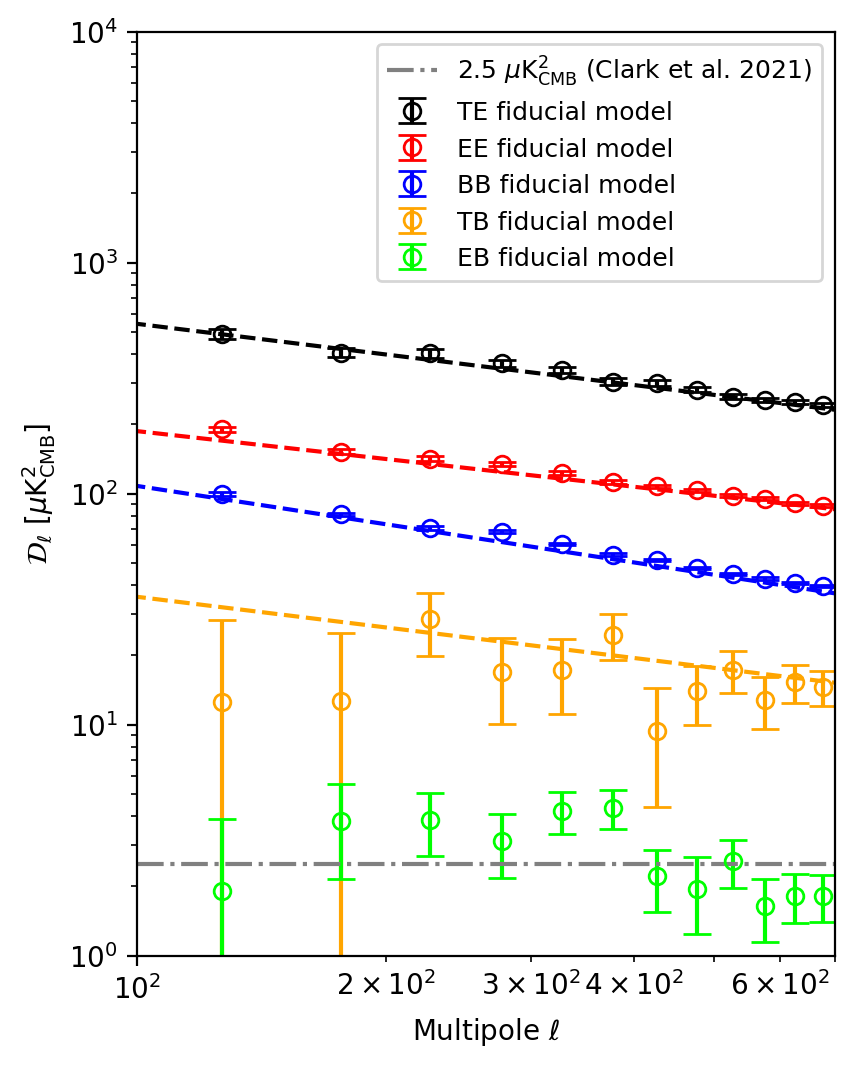}
    \caption{
    Angular power spectra at 353\,GHz in the Galactic plane 70\% mask for our fiducial filament model presented in Sec.~\ref{sec:fiducial_bestfit}. The power-law fit in the same mask to the \textit{Planck} \textsc{npipe} 353\,GHz frequency map (Table~\ref{table:dust_spectra_npipe}) is shown as dashed lines, while the spectra from one realization of our model are the circles in bins with size $\Delta \ell=50$. Error bars are the standard deviation across 100 realizations of the fiducial filament model. The dot-dashed line is $2.5$\,$\mu{\rm K}^2$, the upper limit found by Ref.~\cite{2021ApJ...919...53C}.
    \label{fig:Dell_fiducial_model}
    }
\end{figure}

We produce a realization of our model at higher resolution, $N_{\rm side}=2048$, in the same Galactic plane 70\% mask. We fill in the polarization large scales with the full-mission \textit{Planck} 353\,GHz frequency map filtered in harmonic space such that the overall model fits the dust angular spectra measured (the spectra calculated in Sec.~\ref{sec:planck_data}), following the procedure detailed in Sec.~3.6 of~\citetalias{2022ApJ...928...65H}. This large-scale filling is relevant mostly at scales $\ell < 200$. Fig.~\ref{fig:Dell_fiducial_model} shows the angular power spectra from this realization of our filament model, calculated in the Galactic plane 70\% mask, in the multipole range $\ell=100-700$ in bins with size $\Delta \ell = 50$, for 353\,GHz. 

First, we note that our model matches the $TE$, $EE$, and $BB$ dust spectra measured from \textit{Planck} \textsc{npipe} 353\,GHz, which is shown as power-law fit in dashed lines (the parameters are listed in Table~\ref{table:dust_spectra_npipe}). This of course is by construction, since the filament population in our model is chosen in such a way to fit the dust spectra measured by \textit{Planck}. While the dust $TB$ spectrum is not explicitly used to calibrate the filament model (only the $TE$, $EE$, and $BB$ dust spectra are used), we obtain a good match to the \textit{Planck}-measured dust $TB$ spectrum, being calibrated only from $\hat{\psi}$ estimators measured from observations of \textit{Planck} cross-correlated with HI. Our model also makes a prediction for the $\mathcal{D}_{\ell}^{EB}$ from dust, shown as the green circles. As a reference, we include the signed upper limit prediction from Ref.~\cite{2021ApJ...919...53C} of $\sim 2.5$\,$\mu{\rm K}^2$ for the same mask. We obtain a similar amplitude. 

While the prediction from our model assumes all dust emission is produced within filaments, alternative non-filamentary descriptions such as sheet-like structures produce the same level of $E/B$ asymmetry and $TE$ correlation~\cite{2022A&A...663A.175K}. In future studies, we need to identify alternative mechanisms of parity violation for non-filamentary dust structures, as well as quantify more precisely what fraction of dust emission is filamentary and non-filamentary.

\section{Cosmic birefringence analyses} \label{sec:cosmic_birefringence}

Having presented a model of dust that matches the angular power spectra observed by the 353\,GHz frequency channel of \textit{Planck}, and which also contains an intrinsic non-zero parity-violating signal that is a reasonable match to what we can measure in the sky, in Sec.~\ref{sec:prediction_for_EB} we showed a realization of this model. One realization (or many) can be easily produced and used for analysis of observations and/or producing realistic simulations. In this section, we show a couple of examples of isotropic cosmic birefringence analyses that include our filament model to assess the impact of dust with intrinsic non-zero parity-violating spectra.

We can use our filament model to predict $C_{\ell}^{EB,\rm d}$ in different ways under different assumptions. After introducing the general method to measure cosmic birefringence pioneered by Ref.~\cite{2020PhRvL.125v1301M}, we will show two examples of accounting for foreground parity-violating spectra when constraining an isotropic cosmic birefringence angle.

\subsection{Method} \label{sec:CB_method}

We refer the reader to Refs.~\cite{2020PhRvL.125v1301M,2022PhRvL.128i1302D,2022A&A...662A..10E,2022PhRvD.106f3503E} for specific details on this method. In summary, for a single frequency channel, the observed $EB$ spectrum will take the form~\cite{2019PTEP.2019h3E02M}
\begin{multline} \label{eq:EB_observed}
    C_{\ell}^{EB, \rm o} = \frac{\tan(4\alpha)}{2} (C_{\ell}^{EE, \rm o}-C_{\ell}^{BB, \rm o}) + \frac{C_{\ell}^{EB, \rm fg}}{\cos(4\alpha)} \\
    + \frac{\sin(4 \beta)}{2 \cos(4 \alpha)} (C_{\ell}^{EE, \rm CMB}-C_{\ell}^{BB, \rm CMB}) \text{,}
\end{multline}
where ``o'' means an observed quantity, ``CMB'' and ``fg'' represent the intrinsic spectrum from the CMB and foregrounds, respectively, and $\alpha$ represents the miscalibration angle of the respective channel. Thus the detector angle miscalibration $\alpha$ affects both the CMB and foreground components, while $\beta$ affects the CMB component alone.\footnote{We have omitted a term accounting for a hypothetical intrinsic $EB$ spectrum from the CMB, which is assumed to be zero in the absence of any pre-recombination parity-violating signal, although there are models, e.g.,  Early Dark Energy, that could produce it~\cite{2023PhRvL.131l1001E}.} Eq.~\eqref{eq:EB_observed} can be generalized for multi-frequency observations, accounting for the cross-correlation between channel $i$ and channel $j$~\cite{2020PhRvL.125v1301M}. A Gaussian likelihood that depends on the $\beta$ angle, as well as the calibration $\alpha_i$ angles for each channel $i$, is defined and fitted to the measured cross-spectra (the auto-spectra of each channel are excluded to avoid noise bias) to simultaneously determine the $\beta$ and $\alpha_i$ angles. The intrinsic CMB emission is predicted by computing a theory spectrum using the \textsc{camb} Boltzmann-equation solver~\cite{2000ApJ...538..473L}, using the best-fit cosmological parameters from \textit{Planck} PR3~\cite{2020A&A...641A...6P}, and multiplying by the instrumental beam and pixel window functions to make it directly comparable to the observed spectra.

\begin{figure}
    \includegraphics[width=1.0\columnwidth]{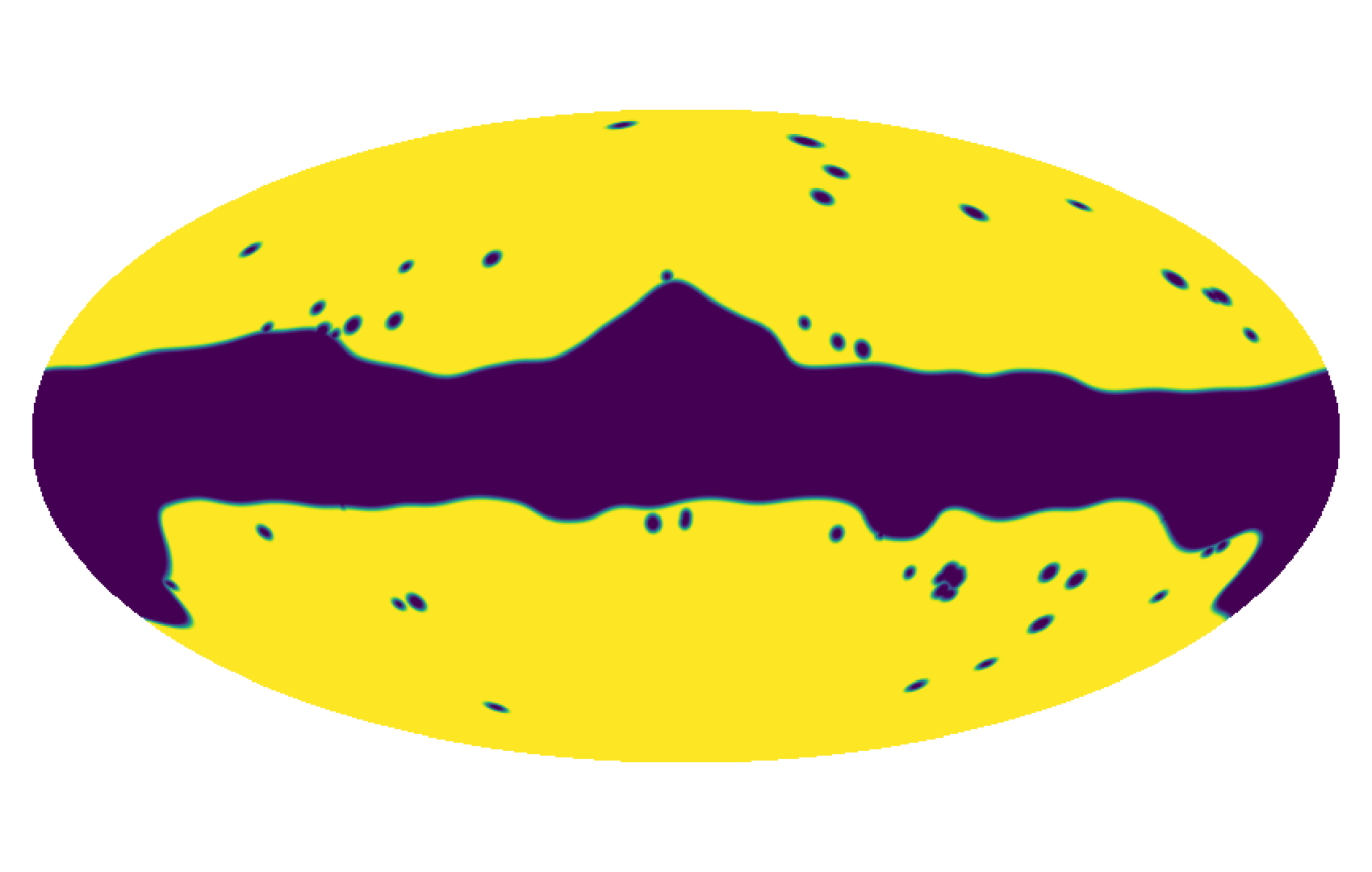}
    \caption{Mask used for our cosmic birefringence analysis. This corresponds to the binary \textit{Planck} Galactic plane 70\% mask joined with the binary mask of strong polarized sources used in the cosmic birefringence analysis of Ref.~\cite{2022PhRvL.128i1302D}. The mask is apodized with a 2$^{\circ}$ scale.  
    }
    \label{fig:gal_70_mask}
\end{figure}

For our run, we will use the \textit{Planck} \textsc{npipe} maps for the HFI frequency channels 100, 143, 217, and 353\,GHz, both A and B detector splits. We will assume dust to be the only significant polarized foreground contribution to eq.~\eqref{eq:EB_observed} at these frequencies as no significant synchrotron $EB$ correlation has been found anyway~\cite{2022JCAP...04..003M,2023MNRAS.519.3383R}, i.e. $C_{\ell}^{EB, \rm fg}=C_{\ell}^{EB, \rm d}$. We will use a mask constructed by using the 70\% Galactic plane (as we have used in this paper so far) binary mask plus the binary mask of the polarized point sources masked in Ref.~\cite{2022PhRvL.128i1302D}. This overall mask is apodized with a 2$^{\circ}$ scale and shown in Fig.~\ref{fig:gal_70_mask}. This mask has $f_{\rm sky}=0.664$. We use \textsc{namaster} to estimate the cross-spectra between channels in the multipole range $\ell\in [51, 1491]$ with $\Delta \ell = 20$. We use $B$-mode purification to calculate all of our angular power spectra to reduce the scatter in the estimation of the pseudo-$C_{\ell}$ and avoid $E$-to-$B$ leakage.

The tightest constraint to date of $\beta \sim \dotdeg{0.34}\pm\dotdeg{0.09}$ quoted in Sec.~\ref{sec:intro} comes from an almost full-sky analysis ($f_{\rm sky}=0.92$)~\cite{2022PhRvD.106f3503E}. As this method uses Galactic emission to calibrate the instrumental angles $\alpha_i$, the larger the sky fraction, the brighter the Galactic emission and the smaller the uncertainties of $\alpha_i$. % Coincidentally, magnetic misalignments seem to average out and have a smaller impact on birefringence measurement for these high $f_{\rm sky}$ masks~\cite{2021ApJ...919...53C}. 
On the contrary, a reduced sky area, such as the $f_{\rm sky}=0.664$ used in this study, will increase the statistical uncertainty of $\beta$. Still, we chose this mask because the higher Galactic latitudes are a cleaner place to isolate the filamentary contribution, losing statistical significance in favor of ensuring a better modeling of dust.

\subsection{Measuring cosmic birefringence from \textit{Planck} HFI data: magnetic misalignment ansatz} \label{sec:CB_misalignment}

The first way of accounting for $C_{\ell}^{EB,\rm d}$ in eq.~\eqref{eq:EB_observed} is to assume the filament-magnetic misalignment ansatz (eqs. \eqref{eq:EB_from_dust}-\eqref{eq:psi_TB_TE}). If we do not know $C_{\ell}^{EB,\rm d}$, we instead can use the $TB/TE$ ratio from dust to estimate the $\psi_{\ell}$ angle, as described in eq.~\eqref{eq:psi_TB_TE} and Ref. \citep{2021ApJ...919...53C}. While our model provides us with $C_{\ell}^{EB,\rm d}$ directly, in this section, we will alternatively use the $\psi_{\ell}$ angle measured from the filament model since it produces more robust estimates of the $TE$ and $TB$ spectra that are calibrated to reproduce all \textit{Planck} dust cross-correlations with HI $\psi$ estimators (the first four estimators of Fig.~\ref{fig:psi_estimators}), as opposed to using a single pure-dust $TB/TE$-measured $\psi_{\ell}$ like previous works have done \cite{2022PhRvL.128i1302D,2022A&A...662A..10E,2022PhRvD.106f3503E}.

We will follow the analysis from Ref.~\cite{2022PhRvD.106f3503E}, whose software is publicly available.\footnote{\url{https://github.com/LilleJohs/Cosmic_Birefringence}} This method implements the use of MCMC with \textsc{emcee}\footnote{\url{https://github.com/dfm/emcee}}~\cite{2013PASP..125..306F} to fit the parameters. In Ref.~\cite{2022PhRvD.106f3503E}, the $\psi_{\ell}$ estimation is done by averaging both combinations of the $TB$ and $TE$ spectra from the A and B split of the 353\,GHz frequency maps. However, as described in Appendix~\ref{sec:systematics_npipe}, the realistic simulations of the \textsc{npipe} processing\footnote{These simulations include beam systematics, gain calibration, bandpass mismatches, and transfer function correction, among others.} show the effect of systematics. Hence, we estimate $\psi_{\ell}$ from the $C_{\ell}^{T_{353 \rm B} \times E_{353 \rm A}}$ and $C_{\ell}^{T_{353 \rm A} \times B_{353 \rm B}}$ spectra, where 353A and 353B label the A and B split of the 353\,GHz frequency map. This approach seems to minimize the bias introduced in the spectra (see Fig.~\ref{fig:npipe_systematics}). We fit four free amplitudes $A_{\ell}$ (eq.~\eqref{eq:EB_from_dust}) in the multipole ranges $51\leq\ell\leq130$, $131\leq\ell\leq210$, $211\leq\ell\leq510$, and $511\leq\ell\leq1491$, following Ref.~\cite{2022PhRvD.106f3503E}. In total, we have 13 free parameters: $\beta$, 8 $\alpha_i$ angles, and 4 $A_{\ell}$ amplitudes.

\begin{table}
    \caption{ 
    Best-fit parameters and their $1\sigma$ uncertainties from the cosmic birefringence analysis using the filament misalignment ansatz. This uses \textit{Planck} \textsc{npipe} HFI frequency channels and the Galactic plane 70\% mask. $\beta$ and $\alpha_i$ angles are in degrees.
    \label{table:cb_results_filaments}
    }
    \begin{ruledtabular}
    \begin{tabular}{l l l}
        \multirow{2}{*}{Parameter} & \multicolumn{2}{c}{How to estimate $\psi_{\ell}$} \\
         & \textsc{npipe} 353 A/B splits & Filament model \\
        \hline
        $\beta$ & $\phantom{-}0.39\pm 0.24$ & $\phantom{-}0.69^{+0.27}_{-0.32}$ \\
        $\alpha_{100 \rm A}$ & $-0.32\pm 0.26$ & $-0.62^{+0.33}_{-0.29}$ \\
        $\alpha_{143 \rm A}$ & $\phantom{-}0.15\pm 0.25$ & $-0.16^{+0.33}_{-0.28}$ \\
        $\alpha_{217 \rm A}$ & $-0.08\pm 0.24$ & $-0.39^{+0.33}_{-0.27}$ \\
        $\alpha_{353 \rm A}$ & $-0.13\pm 0.24$ & $-0.45^{+0.34}_{-0.27}$ \\
        $\alpha_{100 \rm B}$ & $-0.41\pm 0.25$ & $-0.71^{+0.33}_{-0.28}$ \\
        $\alpha_{143\rm B}$ & $\phantom{-}0.09\pm 0.25$ & $-0.22^{+0.33}_{-0.28}$ \\
        $\alpha_{217 \rm B}$ & $-0.13\pm 0.25$ & $-0.44^{+0.33}_{-0.27}$ \\
        $\alpha_{353 \rm B}$ & $-0.08\pm 0.24$ & $-0.39^{+0.34}_{-0.28}$ \\
        $10^2A_{51-130}$ & $\phantom{-}21.6\pm 6.1$ & $\phantom{-}78.0\pm 18.0$ \\
        $10^2A_{131-210}$ & $\phantom{-0}3.7^{+1.3}_{-3.6}$ & $\phantom{-0}8.5^{+2.2}_{-8.4}$\\
        $10^2A_{211-510}$ & $\phantom{-0}4.2^{+1.2}_{-4.1}$ & $\phantom{-}10.7^{+4.7}_{-7.9}$ \\
        $10^2A_{511-1491}$ & $\phantom{-0}6.5^{+2.0}_{-6.3}$ & $\phantom{-}20.3^{+9.2}_{-14.0}$ \\
    \end{tabular}
    \end{ruledtabular}
\end{table}

\begin{figure}
    \includegraphics[width=1.0\columnwidth]{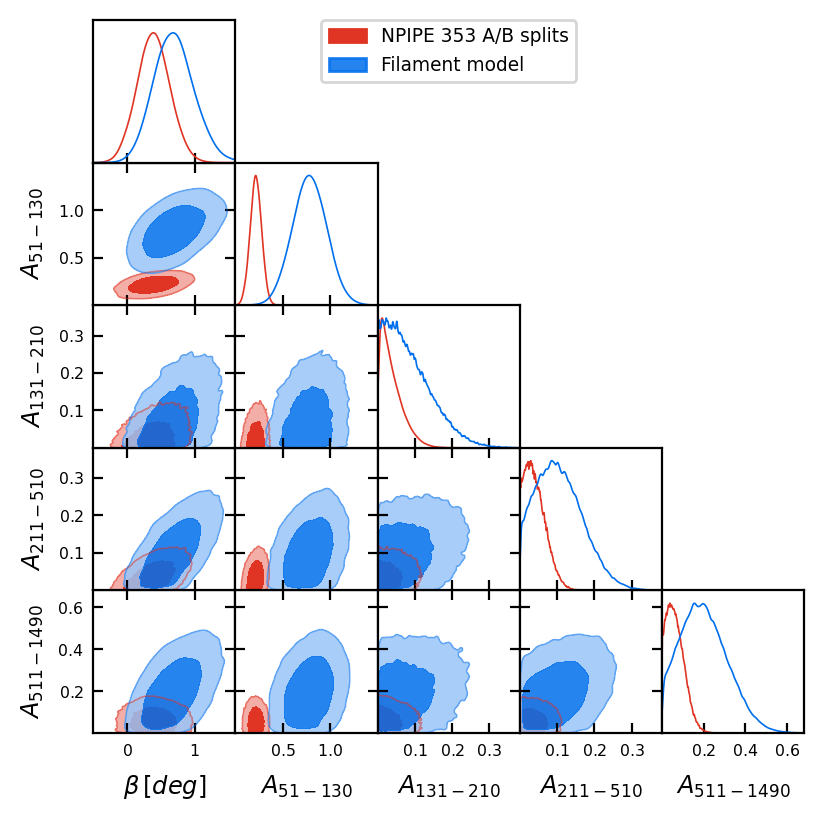}
    \caption{
    Posterior distributions for our cosmic birefringence analysis using the filament misalignment ansatz (eq.~\eqref{eq:psi_TB_TE}), the Galactic plane 70\% mask, and \textit{Planck} \textsc{npipe} HFI frequency channels. We only include the cosmic birefringence angle $\beta$ and the four dust $EB$ amplitudes $A_{\ell}$.
    \label{fig:triangle_plot_realdata_filaments}
    }
\end{figure}

The best-fit parameters and their $1\sigma$ uncertainties for this measurement are listed in Table~\ref{table:cb_results_filaments} and shown in Fig.~\ref{fig:triangle_plot_realdata_filaments} in red. With the 353\,GHz channel split modification to the method of Ref.~\cite{2022PhRvD.106f3503E}, and using \textsc{npipe} data to estimate $\psi_{\ell}$, we measure $\beta=\dotdeg{0.39} \pm \dotdeg{0.24}$. This measurement can be compared to $\beta=\dotdeg{0.29}\pm \dotdeg{0.28}$ measured by Ref.~\cite{2022PhRvL.128i1302D} for the same dataset, multipole range, and method but in a different Galactic mask of similar sky coverage, $f_{\rm sky}=0.63$. As mentioned in Sec.~\ref{sec:CB_method}, the reduced sky fraction limits our SNR. Also, different sky fractions change the effect of dust in the likelihood and somewhat alter the fitted $\beta$, while the larger error bars as $f_{\rm sky}$ decreases make them consistent with each other (see Fig. and Table~1 of Ref.~\cite{2022PhRvL.128i1302D} for fitted $\beta$ versus $f_{\rm sky}$). Considering all of this, our results align with Ref.~\cite{2022PhRvL.128i1302D}.

By contrast, when we estimate $\psi_{\ell}$ from our filament model, we measure $\beta = \dotdeg{0.69}^{\dotdeg{+0.27}}_{\dotdeg{-0.32}}$. We make a high-SNR estimate of $\psi_{\ell}$ by averaging 100 realizations of the fiducial filament model at 353\,GHz. The best-fit parameters and their $1\sigma$ uncertainties are listed in Table~\ref{table:cb_results_filaments} and shown in Fig.~\ref{fig:triangle_plot_realdata_filaments} in blue. The fitted $\beta$ is consistent with estimating $\psi_{\ell}$ from the \textsc{npipe} 353 A/B splits to within $0.83\sigma$, but the $A_{\ell}$ amplitudes change due to the different angular dependence of the $\psi_{\ell}$ estimated from \textsc{npipe} 353 A/B splits and the filament model.

\begin{figure}
    \centering
    \includegraphics[width=1.0\columnwidth]{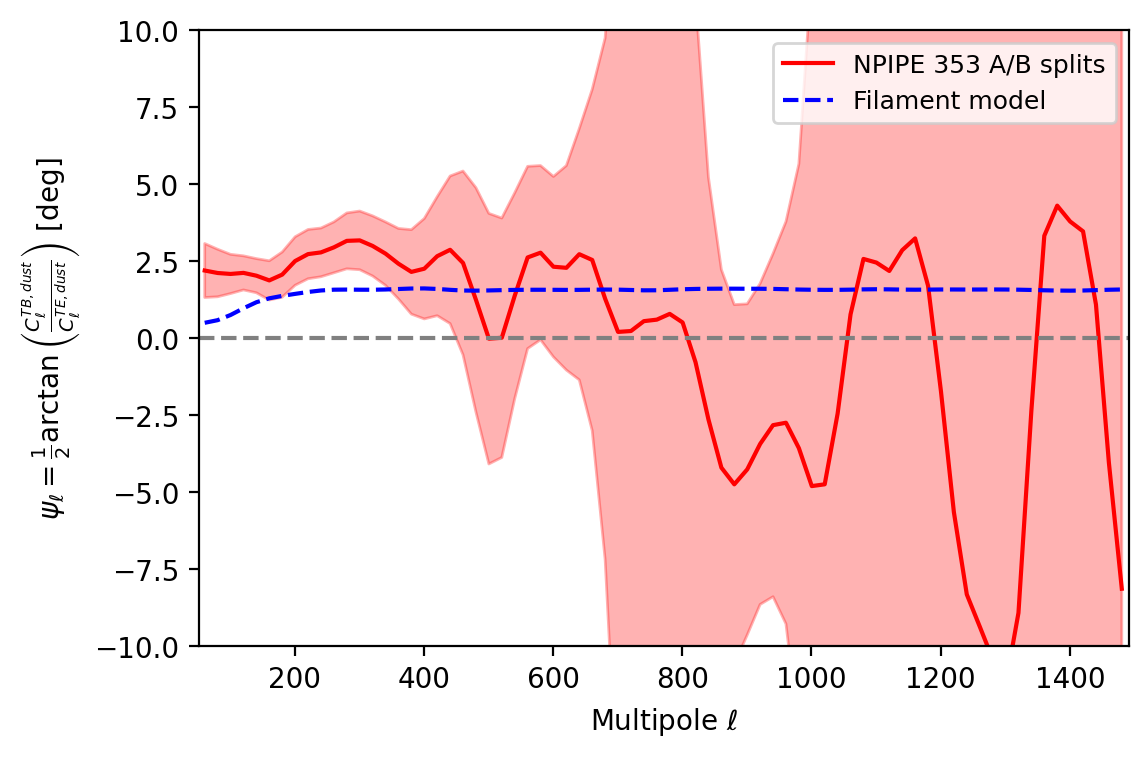}
    \caption{
    $\psi_{\ell}$ estimated from \textsc{npipe} 353 A/B splits in solid red. The red-shaded region represents one standard deviation calculated from 100 \textsc{npipe} simulations. The blue dashed curve is the mean across 100 realizations of our filament model. Note that the uncertainty for the red curve is dominated by the noise in \textsc{npipe} maps.
    }
    \label{fig:psi_ell}
\end{figure}

Fig.~\ref{fig:psi_ell} shows the estimated $\psi_{\ell}$ from \textsc{npipe} 353 A/B splits as described above in solid red. $\psi_{\ell}$ estimated from the mean of 100 realizations of our filament model is in dashed blue. Both include smoothing by a 1D Gaussian filter with $\sigma=1.5$ the width of a bin following Ref.~\cite{2022PhRvD.106f3503E}. Here we can see that using the \textsc{npipe} maps gives $\psi_{\ell} \sim 3^{\circ}$ for large scales $\ell \lesssim 400$, while for smaller scales it oscillates by a large amount, being more consistent with an average $\psi_{\ell} \sim 0^{\circ}$. This explains the $3\sigma$ measurement of $A_{51-130}$, while the other three $A_{\ell}$ are smaller and more consistent with zero. In contrast, the filament model measures a smaller $\psi_{\ell} \sim 2^{\circ}$ value constant for all scales, except for a small dip at $\ell \sim 200$. A smaller angle makes the $\sin(4 \psi_{\ell})$ term in eq.~\eqref{eq:EB_from_dust} smaller so that the $A_{\ell}$ parameters must be larger to compensate. This is clear in Fig.~\ref{fig:triangle_plot_realdata_filaments} and Table~\ref{table:cb_results_filaments}. The $\psi_{\ell}$ estimated from the filament model is mostly scale-independent, which is of course by construction, since we assign $\psi$ angles to filaments randomly without any kind of correlation with filament angular size. We believe this choice is sound since current constraints from data support a scale-independent $\psi_{\ell}$ in the range $\ell=100-700$~\citepalias{2023ApJ...946..106C}. If future data found evidence for significant scale dependence of $\psi_\ell$, this could be incorporated into our model via a distance-dependent misalignment angle.

To understand how the scale dependence of $\psi_\ell$ affects the birefringence measurement, we can approximate the effect of non-zero $C_\ell^{EB,{\rm d}}$ through the effective rotation angle $\gamma_\ell \sim A_\ell \psi_\ell$ as in Refs.~\cite{2020PhRvL.125v1301M,2022PhRvL.128i1302D}. Then, we effectively measure $\beta'=\beta-\langle \gamma_\ell\rangle$ and $\alpha'=\alpha+\langle \gamma_\ell\rangle$ when ignoring the contribution of the dust $EB$ spectrum. We know $C_{\ell}^{EB, \rm d} > 0$ and therefore $\beta' < \beta$ (see Fig.~1 of Ref.~\cite{2022PhRvL.128i1302D} for an illustration of this). A scale-independent $\psi_\ell$ is the most pernicious for birefringence analyses as it leads to a strong degeneracy between $\beta$ and $\gamma_\ell$ in the likelihood and a potential overestimation of the true $\beta = \beta' + \langle\gamma_\ell\rangle$ when correcting for dust $EB$. Hence, we measure a higher $\beta$ with the scale-independent $\psi_\ell$ from the filament model than with the $\psi_\ell$ derived from \textsc{npipe} maps, which averages to zero at high $\ell$.

\subsection{Measuring cosmic birefringence from \textit{Planck} HFI data: using a dust template} \label{sec:CB_template}

The second approach to account for $C_{\ell}^{EB,\rm d}$ in eq.~\eqref{eq:EB_observed} is to measure it directly from a template. This method is detailed in Ref.~\cite{2022PhRvL.128i1302D} and~\cite{2023JCAP...01..044D} (hereafter \citetalias{2023JCAP...01..044D}), where they use the \textsc{npipe} simulations of the \textsc{commander} sky model~\citep[CSM,][]{2020A&A...641A...4P}. We use the~\citetalias{2023JCAP...01..044D} pipeline, which works under the same underlying principles described in Sec.~\ref{sec:CB_method}, but instead of fully sampling the posterior probabilities with MCMC, it follows a maximum-likelihood semi-analytic solution, building a large linear system that iteratively solves for the parameters. The advantage of doing this is speed, converging to a solution in only a few iterations. The covariance of the parameters is estimated using the Fisher information matrix. Appendix B of~\citetalias{2023JCAP...01..044D} demonstrates the equivalency of this method to running a full MCMC sampling.

We use the same \textit{Planck} \textsc{npipe} frequencies and splits as in the previous section, with the same mask and setup, in the first case using the CSM as a template and in the second case using the fiducial filament model dust spectra averaged from $N_{\rm sims}=100$ realizations. A realization of the filament model is produced at 353\,GHz and then extrapolated to the other frequencies with a standard modified blackbody spectral law with spatially variable $\beta_{\rm dust}$ and $T_{\rm dust}$ derived from the \textsc{gnilc} dust maps~\cite{2016A&A...596A.109P}.

\begin{figure}
    \includegraphics[width=1.0\columnwidth]{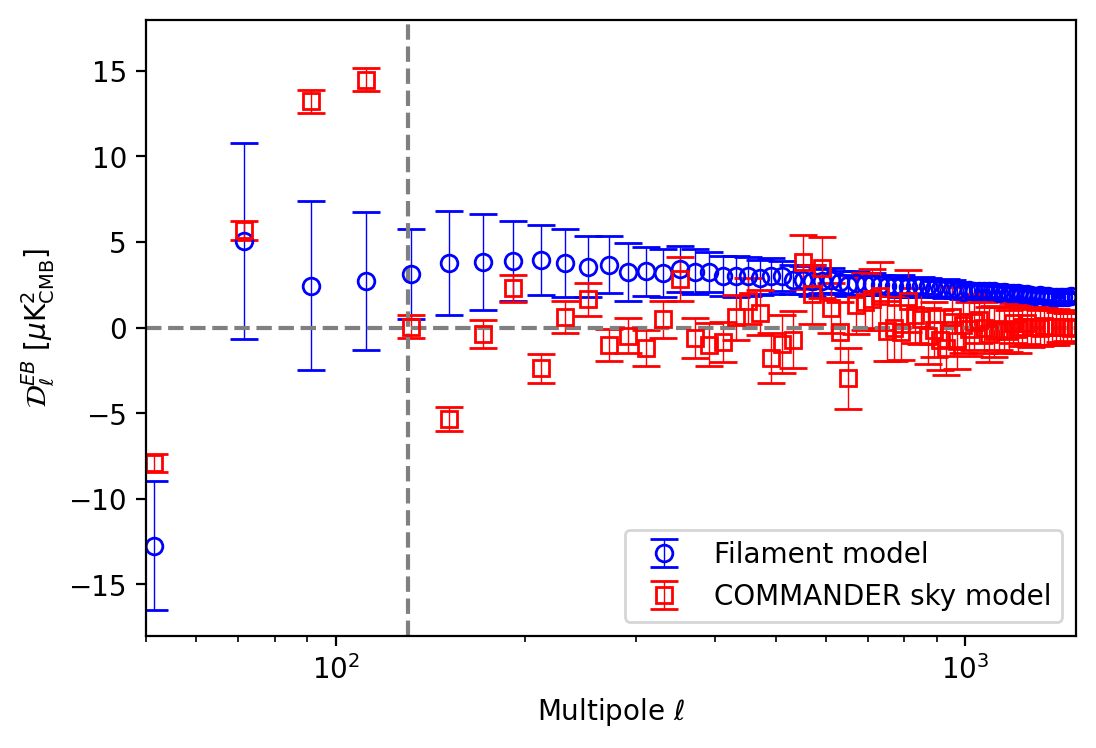}
    \caption{
    $\mathcal{D}_{\ell}^{EB}$ spectrum measured at 353\,GHz in the Galactic plane 70\% mask for our two choices of dust templates: the CSM and the filament model. For the filament model, this is the mean across $N_{\rm sims}=100$ realizations, while the error bars are the standard deviation. For the CSM, the error bars are the standard deviation across 100 realizations of the \textsc{commander} simulations derived with the \textsc{npipe} data. %The maps are extrapolated to 353\,GHz using a constant modified black body with $\beta_{\rm dust}=1.54$ and $T_{\rm dust}=20$\,K. 
    The dashed vertical line marks $\ell=130$.
    \label{fig:template_EB}
    }
\end{figure}

For this analysis, two modifications are made to the~\citetalias{2023JCAP...01..044D} pipeline:
\begin{itemize}
    \item The covariance of eq.~\eqref{eq:EB_observed} includes terms that depend on the dust auto-correlations, $\mathbf{C}^\mathrm{d}_\ell$, and the cross-correlations between dust and the observed data, $\mathbf{C}^\mathrm{d*o}_\ell$. Since the dust signal should only correlate with itself, the analytical expression of $\mathbf{C}^\mathrm{d*o}_\ell$ can be expanded into a combination of the dust $EE$, $BB$, and $EB$ power spectra rotated by the $\alpha_i$ angles (see eq.~D.2 of~\citetalias{2023JCAP...01..044D} for the full expression). The original~\citetalias{2023JCAP...01..044D} considered the CSM as a template of the particular dust emission in our Galaxy and used the CSM auto-spectra and CSM-data cross-spectra to build the covariance. On the other hand, our filament model is an independent realization of a dust model, so its cross-correlation with the \textsc{npipe} frequency maps is null. Nevertheless, we can produce $N_{\rm sims}$ realizations of our model and average them to obtain the approximate underlying fiducial dust spectra to use in the calculation of $\mathbf{C}^\mathrm{d}_\ell$ and $\mathbf{C}^\mathrm{d*o}_\ell$. On a technical level, this means substituting eq.~A.7 of~\citetalias{2023JCAP...01..044D} for
    \small
    \begin{align}\label{eq: fg in cov}
       \mathbf{C}_{ijpq\ell}^{\mathrm{d*o}} = &  - 2{\cal A}_\ell{\rm C}_{ij}{\rm C}_{pq}(C_\ell^{E_iE_p,{\rm d}} C_\ell^{B_jB_q,{\rm d}} + C_\ell^{E_iB_q,{\rm d}} C_\ell^{B_jE_p,{\rm d}})\nonumber\\
        &  - 2{\cal A}_\ell{\rm S}_{ij}{\rm C}_{pq} (C_\ell^{B_iE_p,{\rm d}} C_\ell^{E_jB_q,{\rm d}} + C_\ell^{B_iB_q,{\rm d}} C_\ell^{E_jE_p,{\rm d}})\nonumber\\
        &  - 2{\cal A}_\ell{\rm C}_{ij}{\rm S}_{pq} (C_\ell^{E_iB_p,{\rm d}} C_\ell^{B_jE_q,{\rm d}} + C_\ell^{E_iE_q,{\rm d}} C_\ell^{B_jB_p,{\rm d}})\nonumber\\
        &  - 2{\cal A}_\ell{\rm S}_{ij}{\rm S}_{pq} (C_\ell^{B_iB_p,{\rm d}} C_\ell^{E_jE_q,{\rm d}} + C_\ell^{B_iE_q,{\rm d}} C_\ell^{E_jB_p,{\rm d}}),
    \end{align}
    \normalsize
    where $C_{\ell}^{E_iE_j,{\rm d}}$, $C_{\ell}^{B_iB_j,{\rm d}}$, and $C_{\ell}^{E_iB_j,{\rm d}}$ are the average dust spectra of our model at frequencies $\nu_i$ and $\nu_j$, and
    \begin{align}
       {\rm C}_{xy} = &  \frac{2\cos(2\alpha_x)\cos(2\alpha_y)}{\cos(4\alpha_x)+\cos(4\alpha_y)},\\
       {\rm S}_{xy} = & \frac{2\sin(2\alpha_x)\sin(2\alpha_y)}{\cos(4\alpha_x)+\cos(4\alpha_y)} \text{.}
    \end{align}
    \item The original~\citetalias{2023JCAP...01..044D} pipeline multiplied the dust spectra in eq.~\eqref{eq:EB_observed} by a single ad-hoc amplitude parameter $\mathcal{A}_{\ell}$ for the entire multipole range. In this paper, we fit two dust amplitudes in the multipole range $51 \leq \ell \leq 130$ and $131 \leq \ell \leq 1491$. $\ell=130$ seems to be the angular scale where the CSM $\mathcal{D}_{\ell}^{EB}$ spectrum transitions between being slightly positive to consistent with zero (see Fig.~\ref{fig:template_EB}).
\end{itemize}

\begin{table}
    \caption{ 
    Best fit parameters and their $1\sigma$ uncertainties from the cosmic birefringence analysis using a dust template. This uses \textit{Planck} \textsc{npipe} HFI frequency channels and the Galactic plane 70\% mask. $\beta$ and $\alpha_i$ angles are in degrees.
    \label{table:cb_results_template}
    }
    \begin{ruledtabular}
    \begin{tabular}{l l l}
        Parameter & \textsc{commander} sky model & Filament model \\
        \hline
        $\beta$ & $\phantom{-}0.16\pm0.12$ & $\phantom{-}0.06\pm0.25$ \\
        $\alpha_{100 \rm A}$ & $-0.08\pm0.15$ & $\phantom{-}0.02\pm0.27$\\
        $\alpha_{143 \rm A}$ & $\phantom{-}0.38\pm0.13$ & $\phantom{-}0.48\pm0.26$ \\
        $\alpha_{217 \rm A}$ & $\phantom{-}0.15\pm0.12$ & $\phantom{-}0.26\pm0.27$ \\
        $\alpha_{353 \rm A}$ & $\phantom{-}0.07\pm0.09$ & $\phantom{-}0.20\pm0.27$ \\
        $\alpha_{100 \rm B}$ & $-0.17\pm0.15$ & $-0.07\pm0.27$ \\
        $\alpha_{143\rm B}$ & $\phantom{-}0.32\pm0.13$ & $\phantom{-}0.42\pm0.26$ \\
        $\alpha_{217 \rm B}$ & $\phantom{-}0.11\pm0.12$ & $\phantom{-}0.22\pm0.27$ \\
        $\alpha_{353 \rm B}$ & $\phantom{-}0.13\pm0.10$ & $\phantom{-}0.25\pm0.27$ \\
        $\mathcal{A}_{51-130}$ & $\phantom{-}1.01\pm0.04$ & $\phantom{-}2.49\pm0.85$ \\
        $\mathcal{A}_{131-1491}$ & $\phantom{-}1.34\pm0.16$ & $-0.07\pm0.30$ \\
    \end{tabular}
    \end{ruledtabular}
\end{table}

\begin{figure}
    \includegraphics[width=1.0\columnwidth]{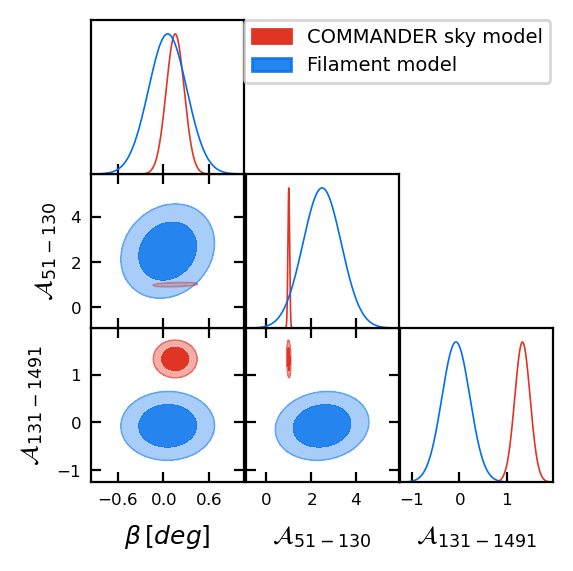}
    \caption{
    Posterior distributions for our cosmic birefringence analysis using a dust template, the Galactic plane 70\% mask, and \textit{Planck} \textsc{npipe} HFI frequency channels. We only include the cosmic birefringence angle $\beta$ and the two dust $\mathcal{A}_{\ell}$ amplitudes.
    \label{fig:triangle_plot_realdata_template}
    }
\end{figure}

The resulting best-fit parameters and their $1\sigma$ uncertainties are listed in Table~\ref{table:cb_results_template} and shown in Fig.~\ref{fig:triangle_plot_realdata_template}. Using the CSM, we measure $\beta=\dotdeg{0.16}\pm\dotdeg{0.12}$, which is comparable to the $\beta=\dotdeg{0.22}\pm\dotdeg{0.18}$ measured by~\citetalias{2023JCAP...01..044D} with the same dataset, multipole range, and method but in a different Galactic mask of similar $f_{\rm sky}=0.63$ and using only a single $\mathcal{A}_{\ell}$ amplitude instead of two.

Using the filament model as a template, we measure $\beta=\dotdeg{0.06}\pm\dotdeg{0.25}$, consistent with the CSM measurement within $0.36\sigma$. %However, there are several things to note. A major factor is that $C_{\ell}^{EB, \rm d}$ in the CSM is more consistent with zero at scales $\ell>130$, while our filament model has slightly smaller values in the large scales and remains roughly constant throughout the entire multipole range of interest. The $\mathcal{D}_{\ell}^{EB}$ spectrum of both templates is shown in Fig.~\ref{fig:template_EB}. Hence, $\mathcal{A}_{51-130}$ is measured to peak at $\sim 2.5$ since a compensation is needed towards higher values of $C_{\ell}^{EB, \rm d}$ to match \textit{Planck}, while $\mathcal{A}_{131-1491}$ is measured to be $\sim 0$ as the \textit{Planck} data seems to average out at $\ell>130$. Given the ${\cal A_\ell}$ values preferred by the filament model, no dust signal will be removed from the covariance matrix as the total dust contribution to the covariance is $\mathbf{C}_{ijpq\ell}^{\mathrm{d}}+\mathbf{C}_{ijpq\ell}^{\mathrm{d*o}}\propto1+{\cal A_\ell}({\cal A_\ell}-2)$~\citepalias{2023JCAP...01..044D}, leaving the filament-model fit with less constraining power. 
However, there are several things to note regarding the modeling of dust in the covariance matrix. As discussed in~\citepalias{2023JCAP...01..044D}, the total contribution of dust to the covariance is $\mathbf{C}_{\ell}^{\mathrm{d}}+\mathbf{C}_{\ell}^{\mathrm{d*o}}\propto1+{\cal A_\ell}({\cal A_\ell}-2)$. Therefore, a good dust model with $\mathcal{A}_\ell\approx 1$ will reduce the covariance and lead to tighter constraints. As we saw in Fig.~\ref{fig:template_EB}, the $C_{\ell}^{EB, \rm d}$ in the CSM is more consistent with zero at scales $\ell>130$, while our filament model has slightly smaller values in the large scales and remains roughly constant throughout the entire multipole range of interest. Hence, $\mathcal{A}_{51-130}$ is measured to peak at $\sim 2.5$ since a compensation is needed towards higher values of $C_{\ell}^{EB, \rm d}$ to match \textit{Planck}, while $\mathcal{A}_{131-1491}$ is measured to be $\sim 0$ as the \textit{Planck} data seems to average out at $\ell>130$. Given the ${\cal A_\ell}$ values preferred by the filament model, no dust signal will be removed from the covariance matrix, leaving the filament-model fit with less constraining power.

Another factor to consider is that we are performing a mode-by-mode fit and subtraction of the dust model from the data in this approach. Thus the CSM allows for better constraints as it is highly correlated with the \textit{Planck} data it was derived from. Nevertheless, as noted in~\citetalias{2023JCAP...01..044D}, using the CSM as a dust model can lead to an over-fitting of ${\cal A_\ell}$ and over-reduction of uncertainties since the template also reproduces some of the noise and fluctuations present in \textit{Planck} data. 

All in all, measuring $\beta$ with the filament model results in a smaller yet still consistent value, although the increased covariance results in an error bar twice as big. We also note that the $\alpha_i$ angles are all consistent across the two different dust templates.

A disadvantage of using a template is that we usually estimate the dust amplitude at one anchor frequency (e.g. 353\,GHz). Then, the template is extrapolated to other frequencies. However, frequency decorrelation changes this picture by creating non-trivial distortions of the dust SED~\cite{2023A&A...672A.146V}. The accuracy of a dust template will therefore be limited by the systematics introduced in the component separation, such as an over-simplistic modified blackbody fitting or the spatial clustering of spectral parameters.
\section{Discussion} \label{sec:discussion}

\subsection{Implications for Galactic dust and magnetic field physics}

One interesting fact of the models summarized in Table~\ref{table:best_fit_models} is that all of them show a roughly constant asymmetry where $\sim 56-62$\% of the filament population has a positive $\psi$ angle. Ref.~\cite{2020ApJ...899...31H} found that $\sim 55$\% of filaments need to have a positive $\psi$ angle to reproduce the \textit{Planck}-measured parity-violating $TB$ spectrum.

We can speculate on the physics of dust filament population asymmetry. For example, we can look at how $\psi$ is estimated for an individual filament. This is given by 
\begin{equation}
    \psi = \atantwo (\hat{\vect{r}} \cdot (\vect{L}_{\perp} \times \vect{H}_{\perp}) , \vect{L}_{\perp} \cdot \vect{H}_{\perp}) \text{,}
\end{equation}
where $\vect{L}_{\perp} = \vect{L} - (\vect{L} \cdot \hat{\vect{r}})\hat{\vect{r}}$ and $\vect{H}_{\perp} = \vect{H} - (\vect{H} \cdot \hat{\vect{r}})\hat{\vect{r}}$ are the projections into the plane of the sky of the filament long semi-axis and local magnetic field, respectively, and $\hat{\vect{r}}$ is the unit vector along the LOS that defines the plane of the sky. For $0 \leq \psi \leq \pi/2$ to hold, we require $\hat{\vect{r}} \cdot (\vect{L}_{\perp} \times \vect{H}_{\perp}) > 0$. Using the definition of the projection and expanding, we find
\begin{equation} \label{eq:r_dot_L_x_H}
    \hat{\vect{r}} \cdot (\vect{L}_{\perp} \times \vect{H}_{\perp}) = \hat{\vect{r}} \cdot (\vect{L} \times \vect{H}) \text{.}
\end{equation}
Enforcing eq.~\eqref{eq:r_dot_L_x_H} to be positive means that for the whole filament population, there must be a slight tendency for this cross-product to point away from the observer. This must hold in opposite LOS's with respect to Earth (e.g. a similar positive $\psi$ angle is measured for the northern and southern Galactic hemispheres).

In our model, all polarized dust emission is due to filaments, and all filaments are drawn from a skewed distribution. We could instead imagine that the filament handedness is imprinted on the polarized sky by only a particular subset of Galactic filaments, e.g., those associated with the most nearby dust. The nearby dust distribution is affected by the presence of the Local Bubble, a cavity surrounding the present-day location of the Sun that was carved out by supernovae \cite[e.g.][]{1987ARA&A..25..303C,2003A&A...411..447L,2018A&A...611L...5A}. We could hypothesize that the Local Bubble is related to the presence of magnetically misaligned dust filaments, i.e., this is a phenomenon of the nearby dust, and more distant filaments contribute no parity-odd signal. We would then need to explain the parity-odd polarized intensity distribution with only the emission from this Local-Bubble-associated dust. Coupled with the fact that Ref.~\cite{2024ApJ...973...54H} detects a contribution to the measured 353\,GHz polarized dust emission from dust beyond the Local Bubble wall, this would lead us to interpret the skewness of our fitted misaligned filament model as a lower limit. However, the possibility that the non-filamentary component of polarized dust could produce parity-violating emission through some unknown mechanism complicates this idea, and in that case, the required skewness might be lower or higher.

Specific conditions in the ISM could explain in the future the non-zero parity violation signal. For example, Ref.~\cite{2024ApJ...972...26S} performs idealized simulations of MHD turbulence, finding $EB$ cross-correlation ratio statistically consistent with zero, but showing a slight tendency towards positive values for high-velocity fluids (with high sonic Mach numbers). However, firm conclusions about the $EB$ correlation are hard to draw. In the future, a systematic simulation study might provide some insight into the conditions that produce parity-violating correlations.

\subsection{Implications for cosmic birefringence} \label{sec:discussion_cosmic_birefringence}

In both of the analyses presented in Sec.~\ref{sec:CB_misalignment} and \ref{sec:CB_template}, we find our best-fit $\beta$ to be consistent with the result when using \textit{Planck} data to estimate the $\psi_{\ell}$ angle or when using the CSM as a template for the parity-violating spectrum of dust. We attribute this consistency to the extra degrees of freedom introduced by the ad hoc amplitude parameters for the $EB$ parity-violating spectrum of dust, which can absorb most of the differences between the dust model and the data. In a sense, the small $\leq 0.83\sigma$ shift on $\beta$ we observe here when using a more complex and harmful dust $EB$ spectrum validates the flexibility of the methodology and the robustness of previous results. Note that the consistency of our model with previous analyses was only tested for large Galactic masks where the filamentary contribution to dust is better isolated at the price of increasing the statistical uncertainty on $\beta$ due to the reduced sky coverage. Nevertheless, our conclusions could still hold for smaller masks as the free amplitude parameters would again partially absorb the impact of dust mismodeling.

Our results also share the slight discrepancy between $\beta$ angles derived from the magnetic misalignment ansatz versus using a dust template, as seen in Ref.~\cite{2022PhRvL.128i1302D}, with the former yielding a higher best-fit $\beta$, and the latter, a lower best-fit $\beta$. To some extent, some of this discrepancy could be attributed to the different ways in which each approach takes into account the correlations between the dust model and the data for the calculation of the covariance matrix. In particular, we remind the reader that, unlike the CSM, our filament model is not correlated with the observations and therefore leads to larger error bars when used as a dust template. However, this does not seem to be the origin for the discrepancy in $\beta$ measurements. Definitive conclusions are difficult to draw directly from the analysis of data where unknowns on the dust modeling, the accuracy of estimators, systematics effect, etc. coexist. Hence a useful exercise and application of our model will be to produce simulations of known $\psi$ and $C_{\ell}^{EB, \rm d}$ to evaluate the validity and compatibility of $\beta$ measurements derived through $\psi$ estimates and dust templates in a controlled environment.
\section{Summary and conclusions} \label{sec:conclusions}

In this analysis, we have proposed a model of millimeter emission from Galactic dust, based on filaments, which has a mechanism to generate non-zero parity-violating spectra. We have calibrated this model using full-sky observations from \textit{Planck} and the HI4PI survey to produce a reasonable fit to the sky. As a demonstration of what the model can do, we apply it to measurements of the isotropic cosmic birefringence angle $\beta$ and assess the impact of the non-zero parity-violating $C_{\ell}^{EB}$ spectrum from dust.

Our model is based on having a preferred handedness in the angle $\psi$, which is the angle between the long axis of a filament and the local magnetic field projected into the plane of the sky. In our filament model, first presented in~\citetalias{2022ApJ...928...65H}, we impose a probability distribution in $\psi$ such that there is an asymmetry between positive and negative angles. In this analysis, we show examples using the Asymmetric Laplace and off-center Normal distributions. To calibrate the required level of asymmetry, we use the $\psi$ estimators defined in~\citetalias{2023ApJ...946..106C}, which use the cross-correlation between millimeter dust observations (e.g., \textit{Planck}) with a HI-derived template using a method that extracts the filaments' orientations from 21-cm spectral data (e.g., the full-sky HI4PI survey). In this work, we explore the use of the Rolling Hough Transform and the Hessian method. A fiducial model with an asymmetry of $\sim 56$\% of filaments having a positive $\psi$ angle is favored by the observations. The power spectra of this model and a prediction for Galactic dust $EB$ emission is presented in Sec.~\ref{sec:prediction_for_EB} and Fig.~\ref{fig:Dell_fiducial_model}. This is consistent with $\mathcal{D}_{\ell}^{EB} \sim {\rm few} \mu{\rm K}^2$, similar to the upper limit of $2.5$\,$\mu$K$^2$ given by Ref.~\cite{2021ApJ...919...53C}. 

When performing a fit of our model, we generate a single realization per parameter set and are subjected to cosmic variance. Ideally, we would generate many realizations and average for each parameter set, but we do not have the capabilities of running the tens of thousands of realizations that this would require. This source of uncertainty is much smaller than the noise from the \textsc{npipe} maps, but we leave its proper estimation for doing inference with our filament model for future work.

We use our filament model to make a new measurement of isotropic cosmic birefringence using the method pioneered by Ref.~\cite{2020PhRvL.125v1301M}, that exploits the local emission of the Galaxy to break the degeneracy between instrumental polarization angles and the true rotation due to cosmic birefringence. In this method, the parity-violating $C_{\ell}^{EB}$ spectrum from foregrounds must be accounted for. We explore two ways of doing this: assuming a filament-magnetic field misalignment ansatz, and using a dust template that directly measures $C_{\ell}^{EB, \rm d}$. We present measurements of $\beta$ using the \textit{Planck} \textsc{npipe} HFI frequency maps with the \textit{Planck} 70\% Galactic plane mask using our fiducial filament model in both these cases. Our measurements of $\beta$ for both these cases are: $\beta=\dotdeg{0.69}^{+\dotdeg{0.27}}_{-\dotdeg{0.32}}$ assuming the filament misalignment model, and $\beta=\dotdeg{0.06}\pm\dotdeg{0.25}$ using the filament model as a dust template. In both cases, these measurements are consistent with previous results using \textit{Planck} data and its derivatives, such as the \textsc{commander} sky model, within $0.83\sigma$. We conclude that using our filament model as an alternative way of accounting for $EB$ emission from dust has minimal impact on the derived $\beta$ angle. We attribute the consistency to the extra degrees of freedom allowed by the ad-hoc amplitude parameters for the $EB$ spectrum of dust. In this sense, we validate the foreground modeling used by previous cosmic birefringence analyses, while also noticing that making strong claims is hard due to the increased uncertainty in our fitted parameters from diverse factors such as reduced sky fractions. To truly account for the impact of parity-violating dust, a systematic forecast with known inputs and outputs must be performed, which we leave for future work.

In this work, we focus on the impact of intrinsic non-zero parity-violating spectra from Galactic dust, but this is not the only intervening factor. The calibration of the detectors' polarization angle, as well as other related instrumental systematics, could also play a major role in biasing a future measurement of cosmic birefringence~\cite{2021JCAP...05..032A,2022JCAP...04..029V,2022JCAP...01..039K,2022JCAP...03..032D,2023JCAP...01..044D,2023JCAP...03..034M}. Ideally, we would want a precise absolute calibration of instrumental polarization angles and great efforts are being made to improve calibration techniques by directly measuring them from artificial sources~\citep[e.g.][]{2015JAI.....450007J,2017JAI.....640008N,2018SPIE10708E..2AN,2020SPIE11453E..2PD,calibration_space_satellite,2024PASP..136k5001R} or even astrophysical ones, such as Tau A (Crab nebula)~\cite[e.g.][]{2020A&A...634A.100A,2024PhRvD.110f3013A}. In the next few years, these efforts should be able to constrain the detector's polarization angle to within $\leq\dotdeg{0.1}$~\cite{2022SPIE12190E..1XC, 2023RScI...94l4502M,2024ApOpt..63.5079M}. However, such a high-precision calibration is challenging to achieve and might not be possible for all instruments. Thus, future experiments will most likely still need to rely on self-calibration to some extent and, in that case, a good understanding of parity-violating dust emission will be key in obtaining precision measurements of both cosmic birefringence and polarization angles.

We envision that the main usefulness of our model would be to assess the impact of parity-violating dust spectra in forecasting a future measurement of cosmic birefringence by upcoming experiments. Ground-based experiments such as BICEP3~\cite{2022SPIE12190E..1XC}, Simons Observatory~\cite{2019JCAP...02..056A,2021JCAP...05..032A}, CMB-S4~\cite{2016arXiv161002743A}, AliCPT-1~\cite{2024JCAP...10..046D}, and satellites such as LiteBIRD~\cite{2023PTEP.2023d2F01L}, are or will be able to attempt measurements of cosmic birefringence soon. Conversely, our model will benefit from future better measurements of dust polarization, for example with the Fred Young Submillimeter Telescope~\cite{2023ApJS..264....7C}. We leave the forecasting of the ability of future surveys to measure isotropic and anisotropic cosmic birefringence and the impact of non-zero parity-violating dust emission for future papers, testing the same methods used in this work, as well as other methodologies~\citep[e.g.][]{2023PhRvD.108h2005J}. 

The \textsc{dustfilaments} code to generate dust filament models is available at \url{https://github.com/chervias/DustFilaments}.

\begin{acknowledgments}
We thank Eiichiro Komatsu, Baptiste Jost, and Lorenzo Barqu\'in-Gonz\'alez for commenting on a draft of this paper. We thank the Parity Violation from Home 2023 online conference\footnote{\url{https://parity.cosmodiscussion.com}} for providing a place for useful discussions that improved this work. 
CHC acknowledges ANID FONDECYT Postdoc Fellowship 3220255 and BASAL CATA  FB210003. KMH acknowledges NSF award 2009870, NASA award 80NSSC23K0466, and DOE award DE-SC0024462. SEC acknowledges NSF award AST-2106607, NASA award 80NSSC23K0972, and support from an Alfred P. Sloan Research Fellowship. 
The Geryon cluster at the Centro de Astro-Ingenieria UC was extensively used for the calculations performed in this paper. ANID BASAL project FB21000, BASAL CATA PFB-06, the Anillo ACT-86, FONDEQUIP AIC-57, and QUIMAL 130008 provided funding for several improvements to the Geryon cluster.
This research used resources of the National Energy Research Scientific Computing Center (NERSC), a Department of Energy Office of Science User Facility using NERSC award HEP-ERCAP-mp107.
This research has made extensive use of \textsc{numpy} \cite{2020Natur.585..357H}, \textsc{scipy} \cite{2020NatMe..17..261V}, \textsc{matplotlib} \cite{2007CSE.....9...90H}, \textsc{namaster} \cite{2019MNRAS.484.4127A}, \textsc{healpy} \cite{2019JOSS....4.1298Z}, \textsc{emcee} \cite{2013PASP..125..306F}, the \textsc{gnu} scientific library~\cite{galassi2018scientific} and \textsc{getdist} \cite{2019arXiv191013970L}.
\end{acknowledgments}

\appendix
\section{RHT study} \label{sec:RHT_study}

\begin{figure}
    \includegraphics[width=1.0\columnwidth]{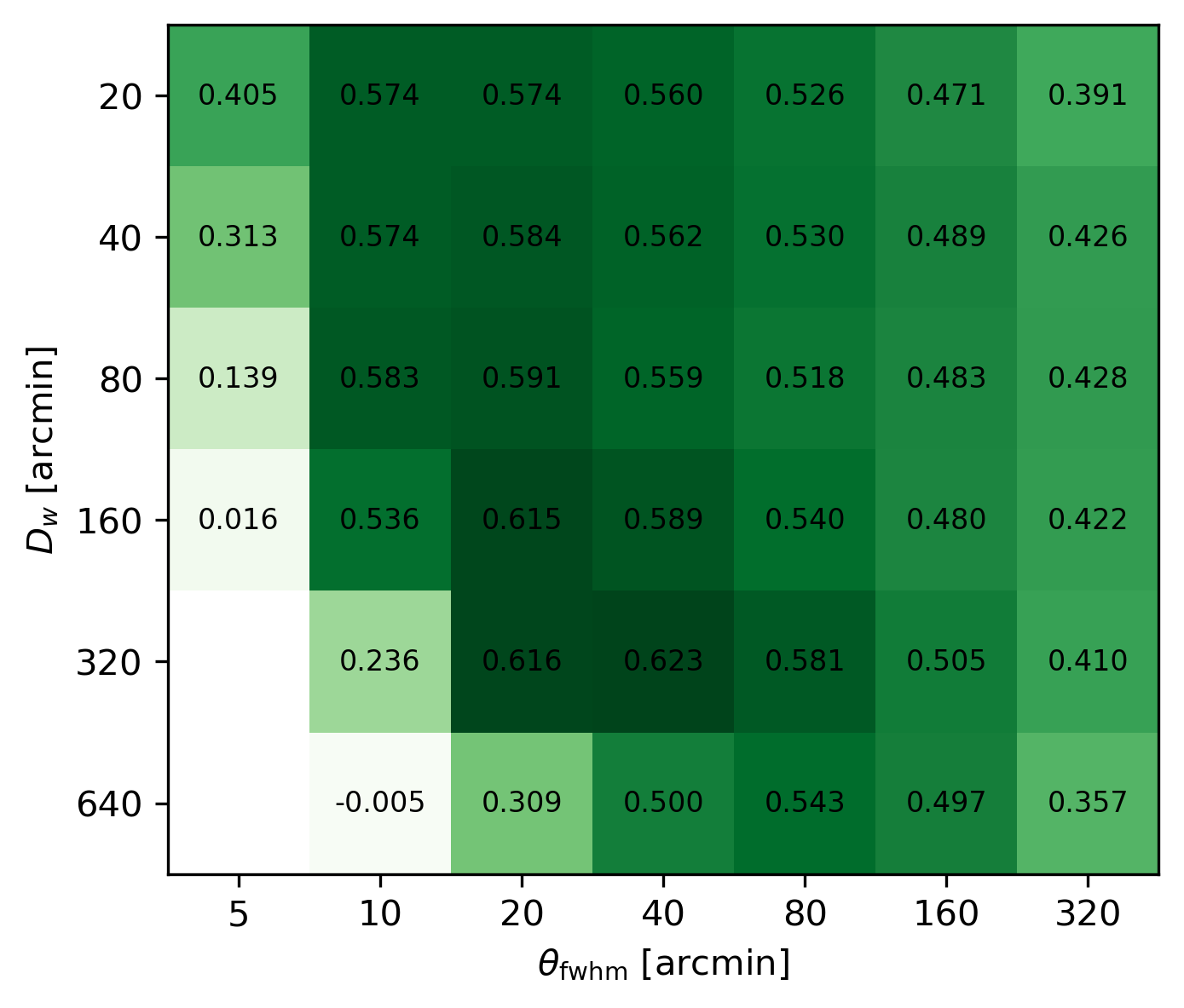}
    \caption{
    Correlation ratio for the $BB$ spectrum between an RHT-derived template and the dust filament model, defined in eq.~\eqref{eq:R_BB}. We use a baseline filament model, using the Galactic plane 70\% mask and a single broad band power in $\ell=20-600$ to estimate the angular spectra. We repeat the procedure for a grid of $(\theta_{\rm FWHM}, D_W)$ parameters.
    }
    \label{fig:sprht_correlation}
\end{figure}

The RHT method, described in Sec.~\ref{sec:rht}, depends on input parameters that can highlight different filament morphologies. Given that we will not include the RHT parameters as parameters in our fit for the best filament model, in this section we describe why we set $Z=0.7$, $\theta_{\rm FWHM}=\dotarcmin{40.0}$, and $D_{W}=\dotarcmin{320.0}$ when constructing a HI-derived template in Sec.~\ref{sec:hessian_vs_rht}.
Following Ref.~\cite{2024ApJ...961...29H}, we fix $Z=0.7$, such that the method is sensitive to filaments larger than 70\% of $D_W$. We construct the same logarithmically-spaced grid of parameters $\theta_{\rm FHWM}$ between $\dotarcmin{5.0}$ and $\dotarcmin{320.0}$, and $D_W$ between $\dotarcmin{20.0}$ and $\dotarcmin{640.0}$. We take the baseline filament dust model, as described in Sec.~\ref{sec:filament_model}, and smooth it to a resolution of $\dotarcmin{16.2}$, the limiting resolution of the HI4PI survey. We do this for both the $TQU$ dust maps, as well as the 20 intensity concentric radial shells. We calculate the RHT over these dust filament maps for the grid of $(\theta_{\rm FWHM}, D_W)$ parameters. We calculate the $BB$ correlation ratio $\mathcal{R}^{BB}$ between the RHT HI-derived dust template and the actual dust template, defined by 
\begin{equation} \label{eq:R_BB}
    \mathcal{R}^{BB} = \frac{C_{\ell}^{B_{\rm d} \times B_{\rm HI}}}{\sqrt{C_{\ell}^{B_{\rm d} \times B_{\rm d}}C_{\ell}^{B_{\rm HI} \times B_{\rm HI}}}} \text{,}
\end{equation}
where the angular power spectra is calculated over the \textit{Planck} Galactic plane 70\% mask in one bandpower bin with multipole range $\ell=20-600$. The $\mathcal{R}^{BB}$ ratios as a function of $\theta_{\rm FWHM}$ and $D_W$ are shown in Fig.~\ref{fig:sprht_correlation}. $\mathcal{R}^{BB}$ is maximized for $\theta_{\rm FWHM}=\dotarcmin{40.0}$ and $D_{W}=\dotarcmin{320.0}$, which we adopt as our parameters to run the RHT over our filament model maps in the main analysis. 

As Ref.~\cite{2024ApJ...961...29H} in their Fig.~12, we find the highest correlation in the range $\theta_{\rm FWHM} \sim \dotarcmin{10.0}-\dotarcmin{40.0}$, which is roughly where the limiting resolution of $\dotarcmin{16.2}$ is. The filaments at these scales are the ones that carry the most information about the underlying magnetic field orientation. The comparison between the RHT reconstruction of $Q$/$U$ and the filament model can be seen in the right-hand side greyscale panels of Fig.~\ref{fig:HI_template}, where the high degree of correlation is evident.

\begin{figure}
    \includegraphics[width=1.0\columnwidth]{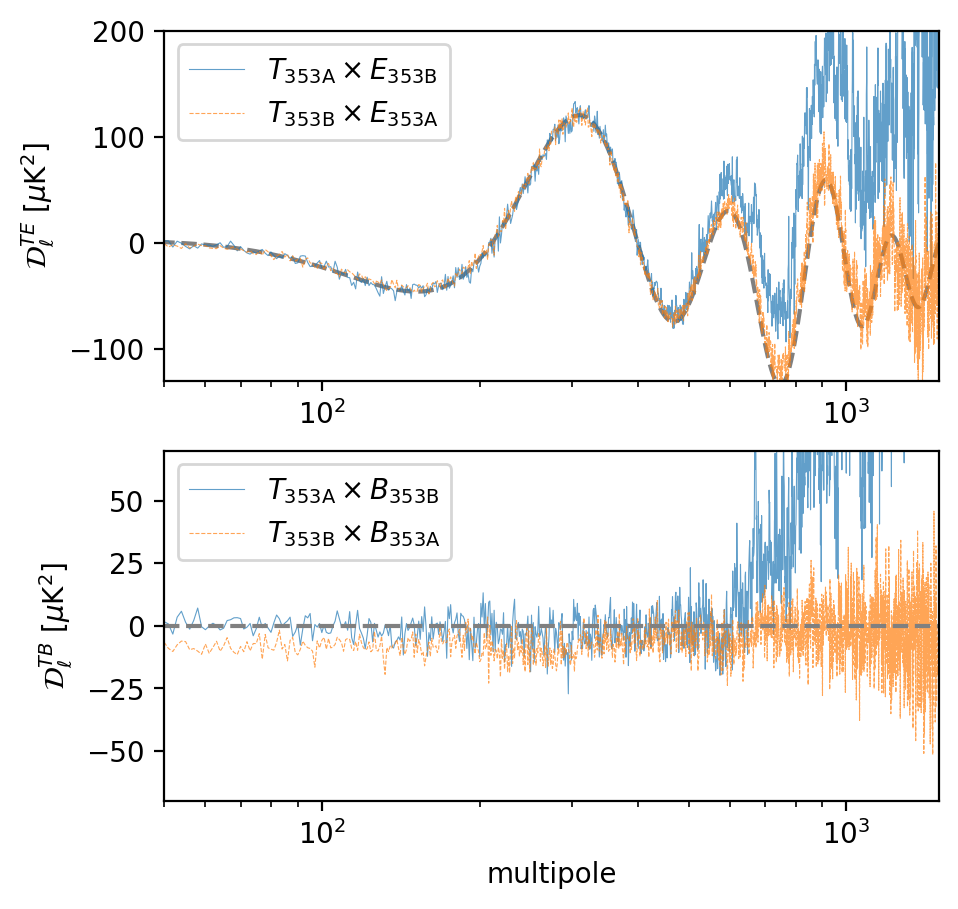}
    \caption{Full-sky $TE$ and $TB$ spectra for both combinations of cross-correlating splits A and B of the 353\,GHz channel. 100 realizations of the end-to-end \textsc{npipe} CMB+noise simulations are averaged. The dashed gray line shows the fiducial CMB spectra.}
    \label{fig:npipe_systematics}
\end{figure}

\section{Systematics in the \textit{Planck} NPIPE data} \label{sec:systematics_npipe}

In Fig.~\ref{fig:npipe_systematics}, we show the average $TE$ and $TB$ full-sky spectra in the range $\ell=50-1491$ and across 100 of the end-to-end \textsc{npipe} 353\,GHz simulations. These simulations correspond to a fiducial CMB realization plus a realistic end-to-end noise realization, for both detector splits A and B. In each case, we show the A$\times$B and B$\times$A cross-correlations separately. Since we have a known fiducial CMB, the average of the 100 realizations should converge to the fiducial CMB $TE$ spectra (shown as the grey line) or to zero in the case of $TB$. However, this is not the case, since the $T_{\rm353 A} \times E_{\rm 353 B}$ spectrum has a clear excess at higher multipoles, while the $T_{\rm 353B} \times B_{\rm 353A}$ spectrum averages to a negative value at low multipoles. Thus, instead of taking an average between A$\times$B and B$\times$A for estimating $\psi_\ell$, we use $T_{\rm 353B} \times E_{\rm 353A}$ for $TE$ and $T_{\rm 353A} \times B_{\rm 353B}$ for $TB$ to avoid these systematics.\footnote{While the $T_{\rm 353A} \times B_{\rm 353B}$ spectrum shows a large bias at $\ell>600$, the signal-to-noise contribution to the overall measurement at these scales is minimal.}

These spurious $TE$ and $TB$ correlations originate from the instrumental systematics (like, e.g., intensity-to-polarization and beam leakage~\cite{Hu:2003dsf, Yadav:2010psy, Miller:2009syn}) that couple and mix the $TT$, $EE$, $BB$, $TE$, $TB$, and $EB$ sky signals of both observed data and simulations. Evidence of such mixing is the significant correlation that exists between dust emission and the CMB+noise simulations used in Fig.~\ref{fig:npipe_systematics} even when dust is not explicitly added to the maps. Even for small leakages, the dust and CMB temperature's relative brightness compared to polarization can lead to appreciable biases in the observed $TE$ and $TB$ correlations. While this effect is very noticeable for full-sky spectra, its impact is significantly reduced when masking with the 70\% Galactic-plane mask as leakages from the dust signal diminish. Nevertheless, our analysis takes the cross-spectra between detector splits that seem more robust against systematics in the full sky.

\bibliography{biblio}

\end{document}